%% file: Main.tex
\documentclass[letterpaper, 10pt]{IEEEtran}

\usepackage[utf8]{inputenc}
\inputencoding{utf8}
\IEEEoverridecommandlockouts
   \usepackage[pdftex]{graphicx}
 
  \graphicspath{{../pdf/}{../jpeg/}}
  \DeclareGraphicsExtensions{.pdf,.jpeg,.png,.jpg}
  
  \usepackage{amssymb} 
\usepackage[natural]{xcolor}

\usepackage{changes}

\usepackage{siunitx}

\usepackage{nicefrac}

\usepackage{bm}

\usepackage[cmex10]{amsmath}

\usepackage{float}
\usepackage{subcaption}

\usepackage{url}

\usepackage{todonotes}

\usepackage{booktabs}
\usepackage{siunitx}

\usepackage{hyperref}
\newcolumntype{N}{>{\centering\arraybackslash}m{1.5cm}}
\begin{document}

\title{\LARGE \bf
  Quantile Surfaces -- \\  \LARGE Generalizing Quantile Regression to Multivariate Targets}


\author{Maarten Bieshaar, Jens Schreiber, Stephan Vogt, Andr\'e Gensler, and Bernhard Sick
  \thanks{M. Bieshaar, J. Schreiber, A. Gensler and B. Sick are with the Intelligent Embedded Systems Lab, University of Kassel,
    Kassel, Germany
      {\tt\footnotesize \{mbieshaar, jens.schreiber, gensler, bsick\}@uni-kassel.de}}
  \thanks{S. Vogt is with the Fraunhofer Institute for Energy Economics and Energy System Technology, Kassel, Germany
      {\tt\footnotesize stephan.vogt@iee.fraunhofer.de}}
}

\maketitle

\input{sections/abstract}



%
\IEEEpeerreviewmaketitle



\input{sections/introduction}
\input{sections/related_work}
\input{sections/method_overview}

\input{sections/evaluation_method}

\input{sections/results_outline}
\input{sections/conclusion}

\section*{\large Acknowledgment}
This work results from the project TRANSFER (01IS20020B) funded by BMBF (German Federal Ministry of Education and Research), KI Data Tooling (19A20001O) funded by BMWI (German Federal Ministry for Economic Affairs and Energy), and the project DeCoInt$^2$ supported by the German Research Foundation (DFG) within the priority program SPP 1835: “Kooperativ interagierende Automobile”, grant number SI 674/11-2.



\bibliographystyle{IEEEtran}
%

{\small
  \bibliography{IEEEabrv,mb,js,sv,ag}
}
\vskip -25mm

\begin{IEEEbiography}[{\includegraphics[width=27mm,clip,keepaspectratio]{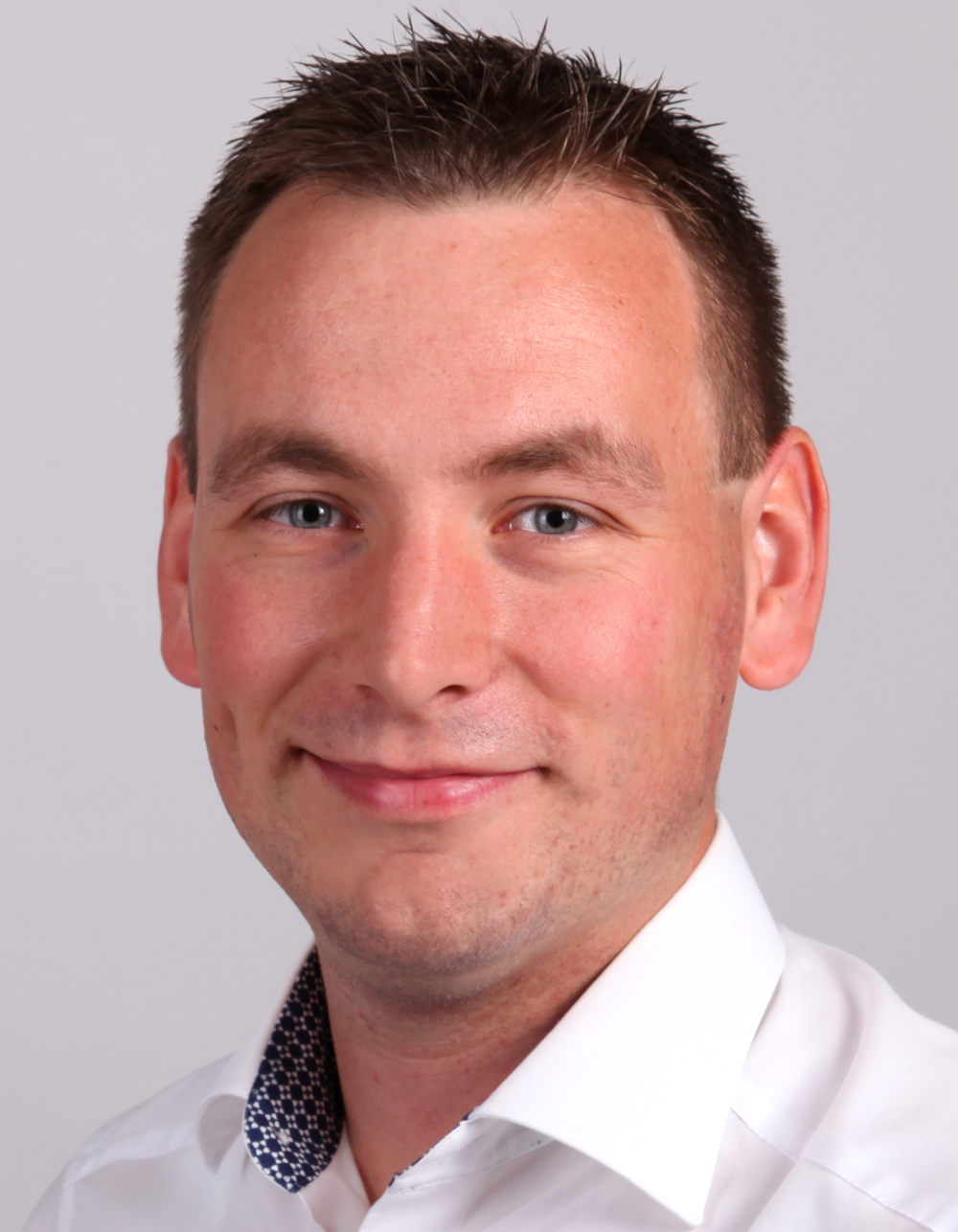}}]
  {Maarten Bieshaar} Maarten Bieshaar received the B.Sc. and the M.Sc. degree in Computer Science from the University of Paderborn, Germany, in 2013 and 2015, respectively. Currently, he is working toward the PhD degree at the University of Kassel, Germany, where he is part of the Intelligent Embedded Systems research group chaired by Bernhard Sick.
\end{IEEEbiography}

\vskip -20mm
\begin{IEEEbiography}[{\includegraphics[width=28mm,clip,keepaspectratio]{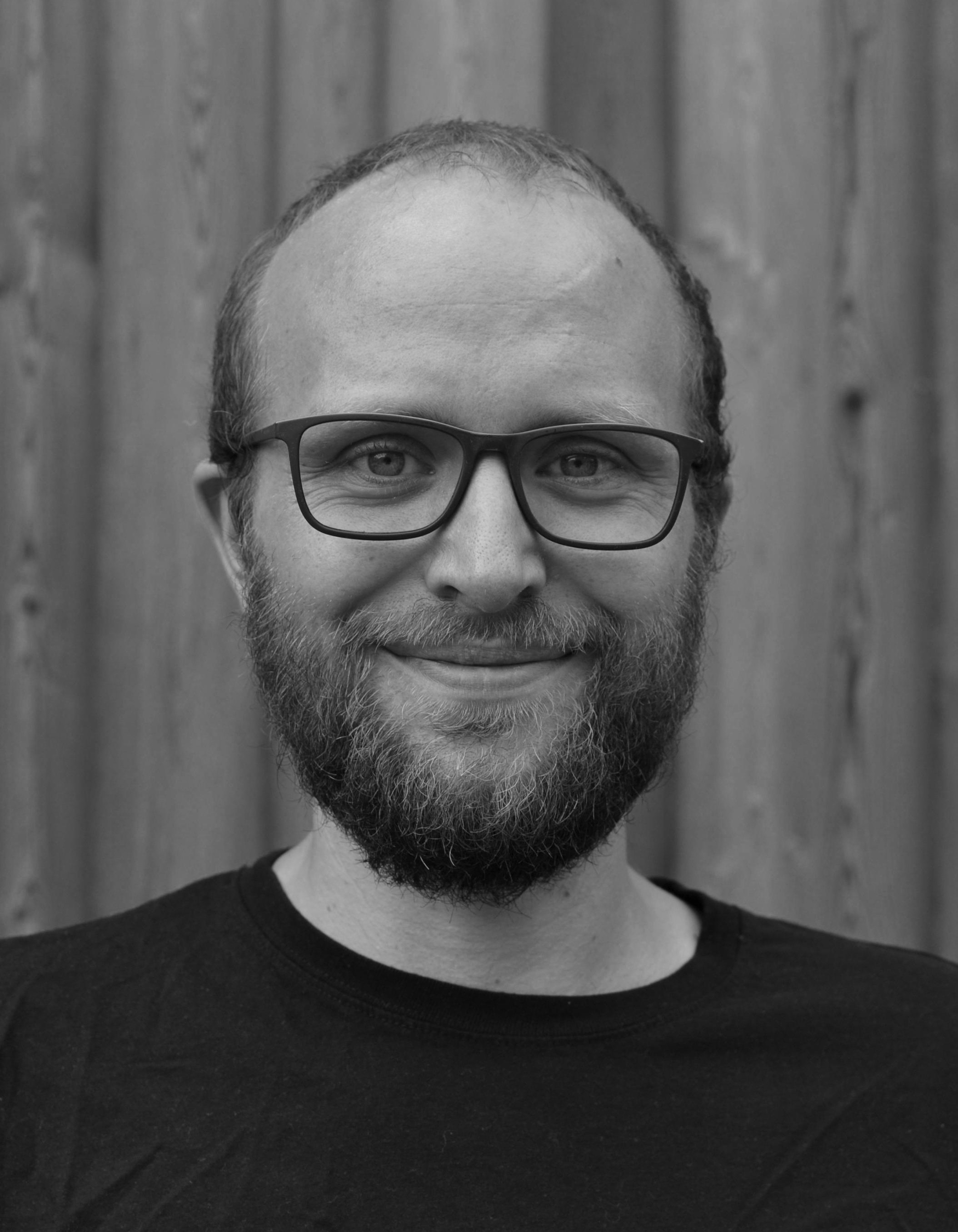}}]
  {Jens Schreiber}  Jens Schreiber received the B.Sc. and the M.Sc. degree in Computer Science from the University of Kassel, Germany, in 2011 and 2013, respectively. Currently, he is working toward the PhD degree at the University of Kassel, Germany, where he is part of the Intelligent Embedded Systems research group chaired by Bernhard Sick.
\end{IEEEbiography}

\newpage

\begin{IEEEbiography}[{\includegraphics[width=27mm,clip,keepaspectratio]{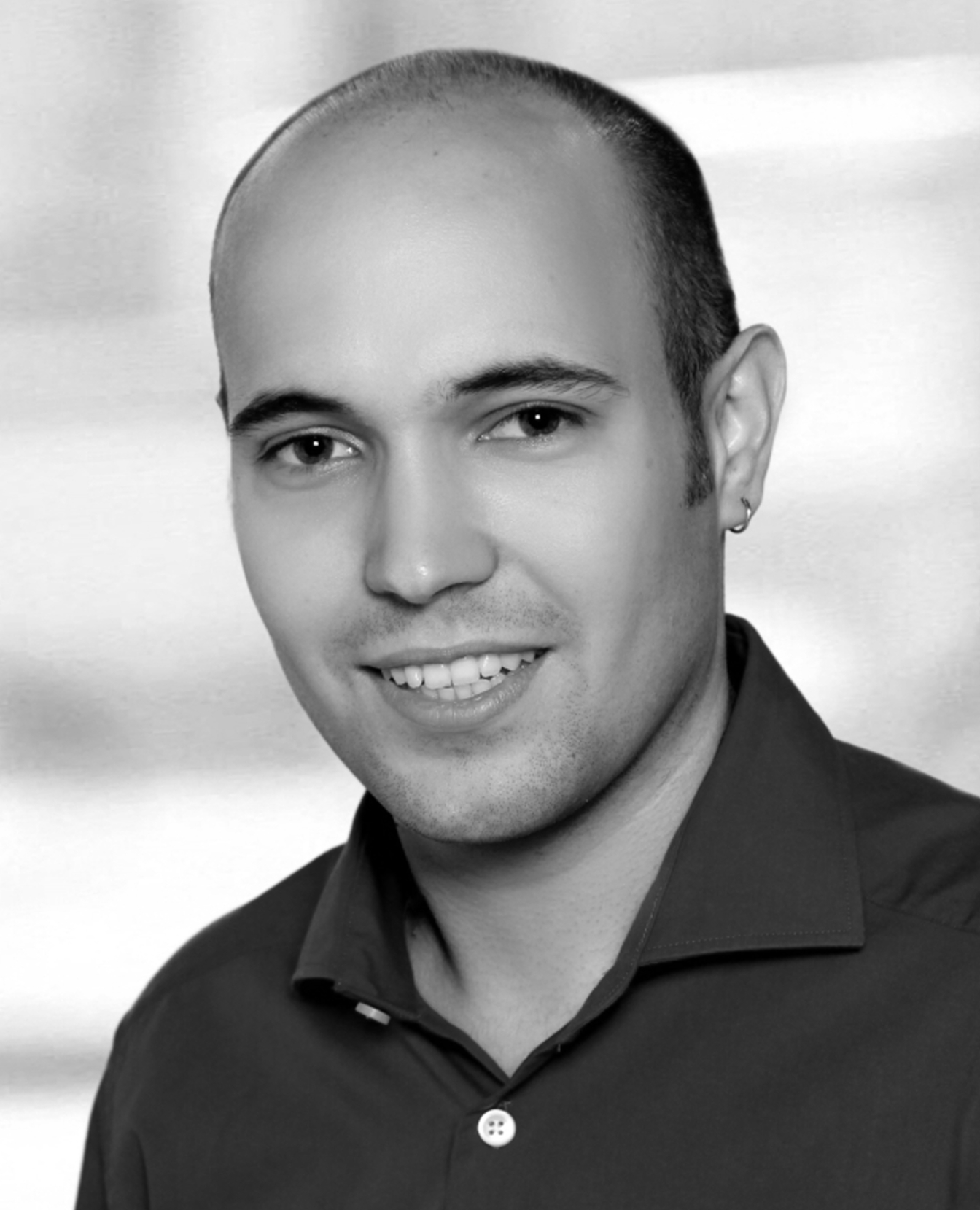}}]
  {Stephan Vogt} Stephan Vogt studied Renewable Energy Technology (B.Eng, FH Nordhausen) and Renewable Energy and Energy Efficiency (M.Sc., University of Kassel).  Since 2014, he is a research associate at the Fraunhofer Institute for Energy Economics and Energy System Technology IEE (formerly Fraunhofer IWES) in Kassel.
  As part of his job, he develops physical and statistical models of wind energy and photovoltaics. He deals with the application of modern methods of machine learning in the field of intraday and day-ahead power forecasting as well as in predictive maintenance for wind turbines. His PhD thesis is furthermore supervised by Bernhard Sick at the University of Kassel.
\end{IEEEbiography}

\vskip -95mm

\begin{IEEEbiography}[{\includegraphics[width=27mm,clip,keepaspectratio]{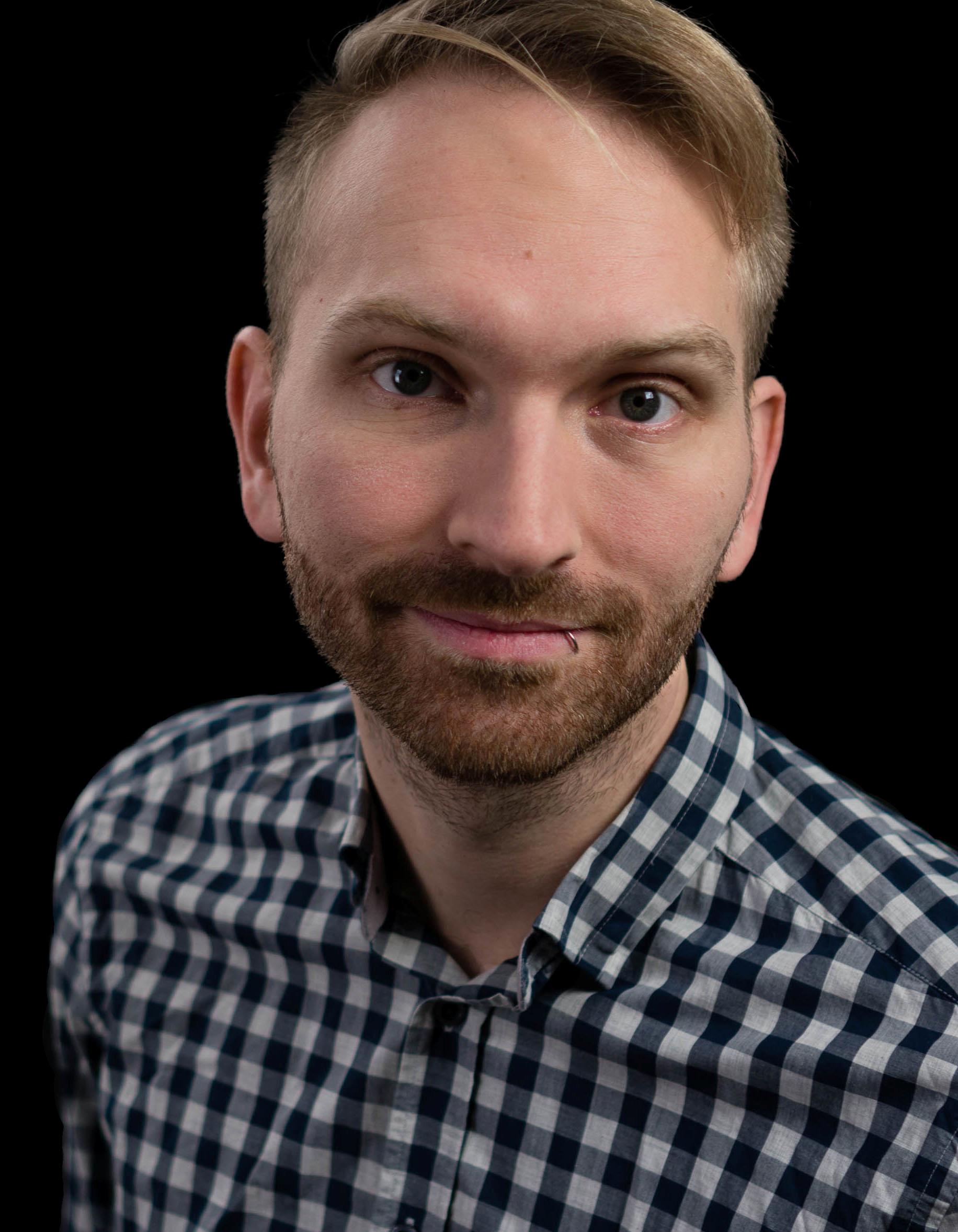}}]
  {Andr\'e Gensler} Andr\'e Gensler got his masters degree in “Electrical Engineering and Information Technology” in 2012 at the University of Applied Sciences in Aschaffenburg,
  Germany before receiving his doctor’s degree in computer science in 2018 at the University of Kassel, Germany. He authored around 20 peer-reviewed
  publications in the machine learning community with focus on forecasting renewable energies. Today, he is working at the SMA Solar
  Technology AG as Senior Algorithms and Data Analysis Development Engineer where he participates in the development and execution of the data
  strategy of SMA. His interests revolve around designing machine learning systems and bringing them into production as data-based services.
\end{IEEEbiography}

\vskip -95mm

\begin{IEEEbiography}[{\includegraphics[width=27mm,clip,keepaspectratio]{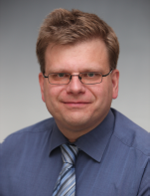}}]
  {Bernhard Sick} Bernhard Sick received the diploma degree in 1992, the PhD degree in 1999, and the ''Habilitation'' degree (university lecturer qualification) in 2004, all in computer science, from the University of Passau, Germany. He is currently a full professor and Intelligent Embedded Systems chair at the Faculty for Electrical Engineering and Computer Science, University of Kassel, Germany. He is an associate editor of the IEEE Transactions on Systems, Man, and Cybernetics-Part B. He holds one patent and received several thesis, best paper, and teaching awards.
  He is a member of the IEEE (Systems, Man, and Cybernetics Society, Computer Society, and Computational  Intelligence Society) and GI (Gesellschaft fuer Informatik).
\end{IEEEbiography}

\end{document}

%% file: sections/abstract.tex
\begin{abstract}
	In this article, we present a novel approach to multivariate probabilistic forecasting.
	Our approach is based on an extension of single-output quantile regression (QR) to multivariate-targets, called quantile surfaces (QS). 
	QS uses a simple yet compelling idea of indexing observations of a probabilistic forecast through direction and vector length to estimate a central tendency.  We extend the single-output QR technique to multivariate probabilistic targets.  QS efficiently models dependencies in multivariate target variables and represents probability distributions through discrete quantile levels. Therefore, we present a novel two-stage process. In the first stage, we perform a deterministic point forecast (i.e., central tendency estimation). Subsequently, we model the prediction uncertainty using QS involving neural networks called quantile surface regression neural networks (QSNN). 
	Additionally, we introduce new methods for efficient and straightforward evaluation of the reliability and sharpness of the issued probabilistic QS predictions. We complement this by the directional extension of the Continuous Ranked Probability Score (CRPS) score. Finally, we evaluate our novel approach on synthetic data and two currently researched real-world challenges in two different domains: First, probabilistic forecasting for renewable energy power generation, second, short-term cyclists trajectory forecasting for autonomously driving vehicles. Especially for the latter, our empirical results show that even a simple one-layer QSNN outperforms traditional parametric multivariate forecasting techniques, thus improving the state-of-the-art performance.

\end{abstract}

%% file: sections/introduction.tex
\section{\large Introduction}
\label{sec_introduction}
\subsection{Motivation}

Forecasting is to predict an uncertain future. 
Probabilistic forecasting takes the form of a predictive probability distribution over possible future events and quantities~\cite{mbGK14}.
It is widely applied in many fields ranging from numerical weather prediction over power forecasting of renewable energies, economics and econometrics, earthquake, and flood prediction up to the prediction of the future trajectory of pedestrians in the automated driving domain. 
Among others, such as mixture-density networks, Bayesian or ensemble techniques, quantile regression (QR) is one of the most popular probabilistic forecasting techniques. 
It is a non-parametric regression technique and does not make any assumptions about the shape of the underlying distribution. 
In contrast to the method of least-squares which aims to estimate a conditional mean, QR aims to determine the conditional quantile (e.g., the median or 50\% quantile) of the response variable~\cite{Bis06}. We obtain the overall predictive distribution by predicting different discrete quantiles levels (e.g., 10\%, 20\%, $\ldots$, 90\%). We can fit a predictive model which issues estimates for particular conditional quantiles by optimization of the pinball loss, i.e., the quantile score. 
We can use various models, ranging from simple linear models over more flexible and expressive non-linear quantile regression neural networks (QRNN)~\cite{Can11} up to gradient boosted regression~\cite{Fri01}.
Nevertheless, besides the expressiveness and flexibility of QR methods, the single-output nature poses a major challenge. 
Since there is no inherent ordering in multi-dimensions, 
extending the notion of quantiles poses a big problem~\cite{Cha03}. Up to now, there is no universally accepted concept of multivariate quantiles~\cite{PS11}.

\begin{figure}
	\centering
	\includegraphics[width=\linewidth, clip, trim=0 0 0 0]{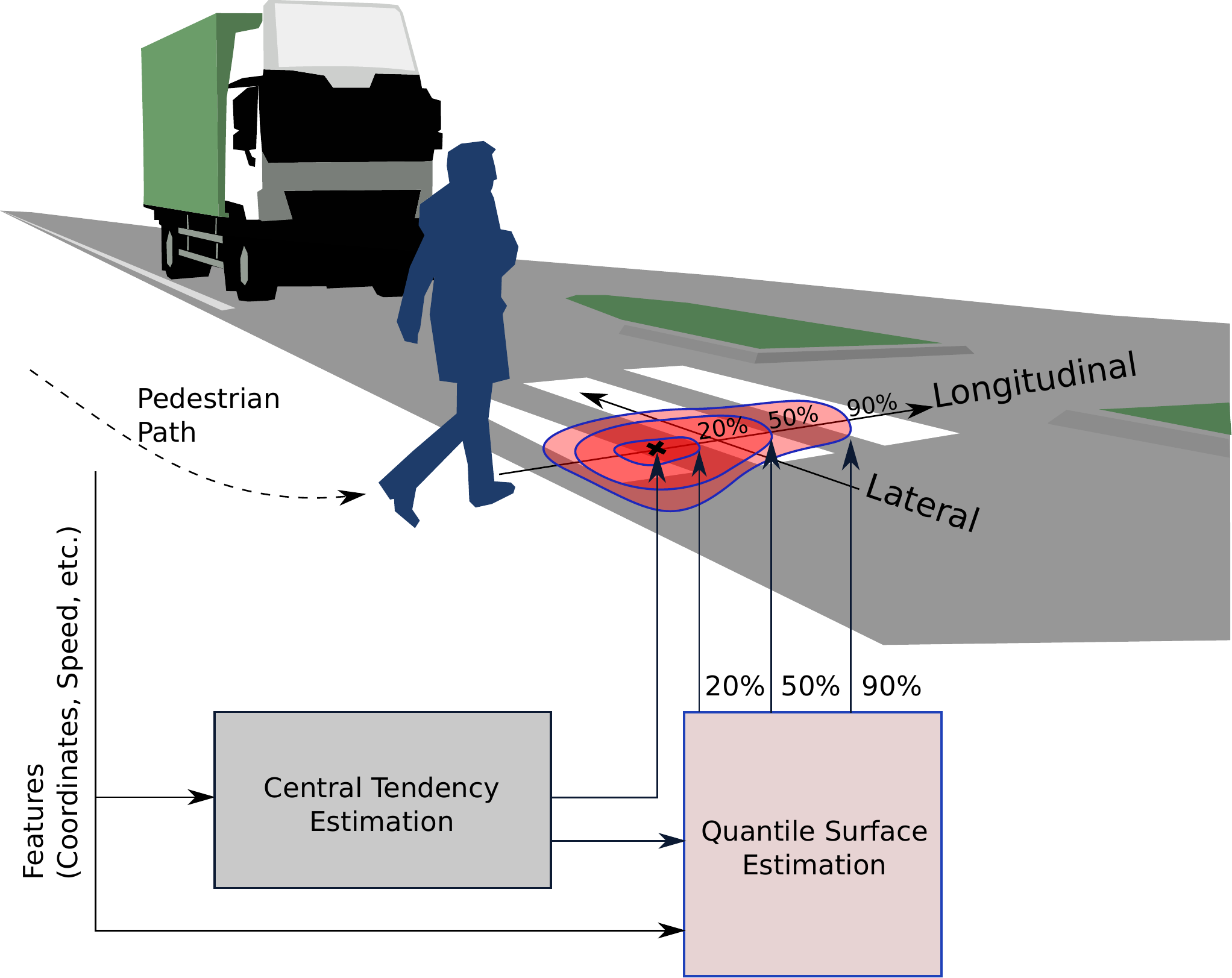}
	\vspace{2mm}
	\caption{Sample application showing the QS forecasting methodology for future pedestrian path prediction (e.g., \SI{1}{\second} ahead). The contour lines around the central tendency estimate (black cross) correspond to different quantile levels and are the quantile surface lines of the QS which are predicted by an artificial neural network forecasting model, the QSNN.}
	\label{fig:sample_qs}
\end{figure}

A natural way of thinking about multivariate quantiles is to think of contour lines around an estimate of central tendency (e.g., mean, mode, or spatial median) describing the underlying conditional distribution. 
This is exemplified in Fig.~\ref{fig:sample_qs} for forecasting a future pedestrian trajectory. 
Following this concept, the contour lines correspond to different quantile levels, whereas the enclosed surface is referred to as \textit{quantile surface (QS)}. Consequently, the contour lines are referred to as quantile surface contour lines.

Assume we are given a deterministic forecast, i.e., an estimate of central tendency, and a set of measured realizations of our multivariate target variable. In the following, we call the latter observations. Our goal is to model the predictive distribution given the estimate of central tendency and the observations.
By indexing observations through direction and distance with respect to the estimate of central tendency, we can recover ordering information. Given a particular direction for an observation, we can now speak of the multivariate observations being close or far away from the estimate of central tendency. 
In related literature, these are also referred to as directional quantiles~\cite{Cha03}.
Using this definition of multivariate quantiles, we can now extend the single-output QR techniques to multivariate output data. 
Assume we are given an estimate of central tendency, we consider indexing of observations utilizing direction and distance to the estimate of central tendency. Here, we interpret the direction as an additional feature entering our forecasting model and the distance, i.e., the vector length, from the central tendency estimate as our target for a particular quantile. 
In doing so, we reduce the problem of estimating multivariate quantiles to predicting one-dimensional quantiles for a vector length depending on the direction as input. This univariate problem can be solved efficiently through traditional techniques such as fitting a standard QR model or a QRNN. Subsequently, the complete quantile contour line of the QS is obtained by sampling the fitted univariate model with different directions. We distinguish between forecasting unconditional probability distributions (i.e., only direction as input feature) and conditional probability distributions. In the latter case, we additionally use other features as input, e.g., in case of pedestrian trajectory forecasting the pedestrian's past trajectory.

Using our method, we can efficiently model the multivariate target variables and represent arbitrary unimodal conditional and unconditional probability distributions utilizing quantiles.

\subsection{Main Contributions and Outline of this Paper}

The key contribution of this article is the extension of QR to multivariate-targets, called quantile surfaces (QS). It is based on a simple, yet effective idea of indexing observations by means of direction and vector length with respect to the estimate of central tendency. Using this methodology, we can issue non-parametric, probabilistic forecasts. Our main contributions can be summarized as follows:

\begin{itemize}
	\item A novel two-stage process for multivariate QR involving neural networks called quantile surface regression neural networks (QSNN) to model star-shaped, unimodal, non-parametric, multivariate, (un-)conditional probability distributions. 
	\item A process for the simple adaptation of arbitrary deterministic central tendency forecasting methodologies to probabilistic forecasting methods using QS.
	\item A novel evaluation metric based on the extension of the continuous ranked probability score (CRPS), allowing to handle the trade-off between sharpness and reliability~\cite{mbGK14} of multivariate probabilistic forecasting techniques.
	\item Investigations involving synthetic and real-world data showing the abilities of the 
	QSs to efficiently model dependencies in multivariate-targets. 
	\item A publicly available implementation of our QS approach using PyTorch and supplementary material\footnote{\url{https://git.ies.uni-kassel.de/public_code/quantile_surface}}.
\end{itemize}

The remainder of this article is structured as follows. 
In Section~\ref{sec_relatedwork}, we review related work in the field of multivariate probabilistic forecasting with special focus on multivariate quantiles. In Section~\ref{sec_method_fundamentals}, we shortly review the fundamentals of probabilistic forecasting.
Then, in Section~\ref{sec_method_overview}, we
introduce the two-stage QS approach and outline its main principles. 
In Section~\ref{sec:evaluation_method}, we present the evaluation methodology for assessing reliability and sharpness as well as the skill~\cite{GSG+08} through the new directional CRPS score. Subsequently, we present the evaluation results on synthetic and two real-world datasets, i.e., short-term cyclist's trajectory prediction and wind power forecasting. Finally, in Section~\ref{sec_conclusion}, the main conclusion and open issues for future work are discussed.

%
%
%
%
%
%
%
%
%
%

%% file: sections/related_work.tex
\section{Related Work}
\label{sec_relatedwork}

In many forecasting fields, the target attributes are inherently multivariate and exhibit mutual dependencies. 
Modeling these dependencies in a conditional distribution forecast is a challenge for probabilistic forecasting
techniques. Arguably, the most well known probabilistic forecasting involving inherently multivariate targets are weather forecast. In weather forecasting, ensemble methods~\cite{RZS16} are widely applied for issuing probabilistic weather forecasts.
In~\cite{LP08}, the authors created probabilistic weather forecasts by re-running the predictive model several times with slightly changing the initial conditions and model parameters.
Other related techniques are Bayesian approaches, e.g., methods based on Markov chain Monte Carlo~\cite{BGH+95} or dropout simulation in Bayesian deep learning~\cite{KG17}.
For both cases, the resulting conditional distribution is represented by samples. 
It can be cast into a fully distributional forecast utilizing kernel density estimation applied as post-processing.
This representation is extremely flexible and allows for modeling of various output distributions.
Still, obtaining the kernel density estimate is computationally expensive, and it is challenging to assess distinct quantiles. Restricting the output to parametric probability distributions, e.g., Gaussian or mixture distributions, can alleviate this shortcoming at the cost of reduced expressiveness. 
A prominent example for these are mixture density networks (MDN)~\cite{Bis06} which are currently regaining attraction 
due to their ability to model arbitrary parametric distributions using artificial neural networks (ANN).
Still, its parametric nature restricts the ability to model arbitrary conditional density functions, and it requires expert knowledge about the shape of the conditional probability distribution to choose the optimal parametrization.
Approaches based on copulas~\cite{TPM15} and recursive Bayesian filters~\cite{TBF05} are also widely applied to multivariate probabilistic forecasting.

QR, as introduced by Koenker and Basset~\cite{KB78}, is nowadays applied to many quantitative fields of research such as economics and econometrics, biomedical studies and clinical trials, biostatistics, power forecasting in the renewable energy domain, and other environmental studies. 
Generalizing the concept of QR to multivariate targets poses a major challenge~\cite{Koe05}. 
The difficulty in extending the notion of QR lies in the lack of an objective basis for ordering multivariate observations. Therefore, there is no generally accepted definition of multivariate quantiles~\cite{PS11}. 
Hence, introducing a (partial) ordering to multivariate data offers an elegant way to tackle this challenge successfully. 

In~\cite{Bar76}, Barnett suggests a different approach to order multivariate data. 
His work is based on the observation that multivariate data also contain some kind of order, e.g., reflecting the extremeness, contiguity, or variability of sample points. Based on the assumption that there is no total ordering in multivariate data, he proposes four different sub-ordering principles that allow us to order given data and look for extreme values and outliers. Still, the article lacks a concept of multivariate quantiles. 
In~\cite{Cha96}, Chaudhuri builds upon this idea introducing a multivariate extension of quantiles based on geometrical considerations. These~\textit{geometrical quantiles} are based on a multivariate extension of the median~\cite{Oja13}, combined with the indexing of multivariate quantiles using directional information, i.e., directional indexing of $K$-dimensional multivariate quantiles by elements of the open unit sphere.
Using this directional information, one can identify ``high 
points'' and ``low points'' with respect to the multivariate dataset.
Based on this indexing scheme, Chaudhuri presents an approach to estimate multi-output linear models to predict the geometrical quantiles. 

In~\cite{KM12}, Kong and Mizera introduce a related concept of \textit{projection quantiles}. They consider the projection of the multivariate targets onto different directional vectors. 
QR is then applied to the length of the different vectors to estimate the location vector of a hyperplane for a particular direction and quantile level. Subsequently, the authors consider the envelope over the hyperplanes of different directions, i.e., Tukey halfspace envelope, for a particular quantile level. The authors show that for elliptical distributions this envelope corresponds to the density contours, 
giving the envelopes a probabilistic interpretation.
The authors' approach can represent arbitrary multivariate conditional distributions by sampling different quantiles and directions using a linear model. 
The authors focus on the theoretical foundations and a rather straightforward application scenario for modeling the body-weight index.
In~\cite{PS11}, Paindaveine and Siman propose a slightly different definition, including a parameterized and piecewise linear function to represent the Tukey halfspace envelope, leading to faster and efficient computation. 

Both, geometrical and projection quantiles are directly related to our definition of QS. Although there is a large variety of research in the field of multivariate geometrical and projection quantiles, as far as our knowledge, this research is mainly focused on theoretical aspects involving simple linear regression models. Besides, our approach does not require a computationally complex calculation of the Tukey halfspace envelope but uses a sample-based representation of the predictive distribution (QS) instead. This can be calculated more efficiently. Moreover, our approach is applicable to large scale real-world problems, which we show in two selected sample applications.

Other related research is the high-dimensional quantile estimation~\cite{SS01}, in which the support of a high-dimensional distribution is modeled in a non-parametric manner using a one-class support-vector machine (SVM). The approach is widely used for outlier detection. The underlying model of the one-class SVM approach is similar to the modeling of multivariate quantiles underlying our QS approach. Nevertheless, the scope (outlier detection against multivariate probabilistic regression) and the algorithmic approach to represent these quantiles differ. 

%% file: sections/method_overview.tex
\section{\large Fundamentals}
\label{sec_method_fundamentals}

Probabilistic forecasting quantifies the uncertainty of a prediction ~\cite{mbGK14}.
In contrast to deterministic forecasting in which one aims at predicting a single value (i.e., point estimate), 
probabilistic forecasting tries to additionally assess the uncertainty of a prediction as given by the data.
If the process underlying the variable is deterministic, a deterministic point estimate is favorable. 
However, if the process is partially stochastic or if we do not have enough data or knowledge, a single point estimate is no longer appropriate since it neglects the uncertainty due to the stochasticity of the process or the lack of knowledge.
Probabilistic forecasting techniques aim to quantify this uncertainty. Besides others, e.g., Fuzzy logic or possibility theory, uncertainty is mostly quantified using probability theory~\cite{LCF13}. 
Our approach also relies on modeling uncertainty using a probability distribution. In the following, this distribution is referred to as \textit{predictive distribution}. 
Ideally, this predictive distribution should reflect the uncertainty or the lack of knowledge of the underlying stochastic process.
According to the origin of uncertainty, we can distinguish \textit{aleatory} and \textit{epistemic} uncertainty. Aleatory uncertainty refers to the natural variability of the physical world, whereas epistemic uncertainty refers to the lack of human or model knowledge. 
The latter can up to a certain level be reduced by means of adding more data, whereas the aleatory uncertainty is inherent and cannot be reduced. 
The QS presented in this article aims to model aleatory uncertainty. We do not explicitly consider epistemic uncertainty.

Probabilistic forecasts are issued for a distinct target space, which is also referred to as \textit{prediction space}. 
Depending on the type of observation, this space can be binary, categorical, or continuous. 
Here, we focus on multivariate continuous cases, i.e., we consider $K$-dimensional real-valued observations $\textbf{o} \in \mathbb{R}^K$. 
Our goal is to issue forecasts for observations based on other input measurements or features, e.g., the past trajectory of a pedestrian. We are interested in issuing a predictive distribution $\hat{p} \left( \mathbf{y} | \mathbf{x} \right)$. This distribution is typically conditional, i.e.,
the probability distribution over the multivariate target $\mathbf{y} := \left( y_0, \ldots, y_K\right)^\text{T}, \mathbf{y} \in \mathbb{R}^K$ is dependent on the $M$-dimensional input feature vector $\mathbf{x} := \left(x_0, \ldots, x_M \right)^\text{T}$. The feature vector may comprise different types of features, i.e., binary, categorical, as well as continuous ones. 
In general, the predictive distribution takes the form of a probability density function (PDF), having the properties

\begin{align}
	\int \hat{p} \left( \mathbf{y} | \mathbf{x} \right) d\mathbf{y} &=1,&  &\hat{p} \left( \mathbf{y} | \mathbf{x} \right) \geq 0.
\end{align}

This predictive distribution can be (approximately) represented in various ways: parametric density functions or mixtures of them~\cite{Bis06}, Monte Carlo simulations~\cite{BGH+95} and ensemble methods~\cite{GSG+08}, prediction intervals~\cite{KNC+11}, and in the univariate case also using QR~\cite{Koe05}. 
QR allows representing the predictive distribution in a non-parametric way using discrete levels of quantiles. 
Moreover, QR provides a decision-theoretical framework for making optimal decisions under uncertainty~\cite{mbGK14}.
In contrast to sample-based representation, quantiles are also well-defined with respect to extreme values of the predictive distribution. 
Notably, the latter is appealing for forecasting in safety-critical applications, e.g., pedestrian path prediction for automated vehicles.
Henceforth, we aim at representing the predictive distribution employing a multivariate generalization of quantiles. 

In the following, we consider the difficulties of generalizing quantiles to multiple dimensions.


\textbf{One-Dimensional Case:}
Consider a real-valued random variable $x \in \mathbb{R}$ with realization $x'$ for which $p \left( x \right)$ denotes the PDF.
The cumulative distribution function (CDF) of $x$ is given by
\begin{align}
P \left( x \right) =  \int_{-\infty}^{x} p \left( t \right) dt.
\end{align} 
Quantiles are ``measures of location'' of a probability distribution. 
They can be understood as a threshold which divides the distribution 
in a given ratio. This ratio describes how many samples are expected 
to be below the threshold. 
A real-valued number $x_\tau \in \mathbb{R}$ is called a $\tau$ quantile if the following holds
\begin{align} \label{eq:quantile}
P \left( x_\tau \right) \geq \tau.
\end{align}
Quantiles are related to the CDF, i.e., quantiles can be directly obtained from the inverse of a CDF. One can use the CDF to derive enclosing boundaries (e.g., $5$\% to $95$\% quantile) which describe the expected location of the data, e.g., $90$\%.

\begin{figure}[ht]
	\centering
	\includegraphics[width=0.9\linewidth]{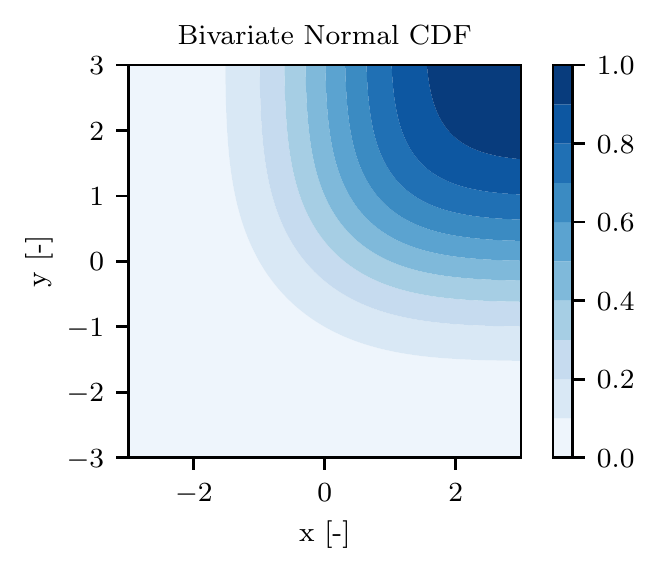}
	\vskip 4mm
	\caption{Schematic of CDF of a bivariate normal distribution with mean $\bm{\mu} = \mathbf{0}$ and $\mathbf{\Sigma} = \sqrt{2}\,\mathbf{I}$.}
	\label{fig:multivariante_cdf}
\end{figure}

\textbf{Multivariate Case:}
We now consider the multivariate extension of quantiles. 
Since there exists no ordering in multiple dimension, we cannot 
extend the notion of quantiles directly~\cite{Cha03}.
For illustration, we study the two-dimensional case for two random variables $x, y \in \mathbb{R}$, where we define for the two real-valued numbers $x_\tau, y_\tau \in \mathbb{R}$ the $\tau$ quantile as follows
\begin{align}
P \left( x_\tau, y_\tau \right) \geq \tau.
\end{align}
Fig.~\ref{fig:multivariante_cdf} depicts the CDF of a two-dimensional normal distribution.
The quantiles are no longer uniquely defined. There are multiple solutions possible, which all lie on a hypersurface. Every point on this surface forms a quadrant, which encloses a region of probability $\tau$.
A quadrant may have a probability mass of $\tau$, although the point on the quantile hypersurface can be arbitrarily far away from the data. 
Hence, due to the lack of any distinct ordering, we cannot merely extend the definition of quantiles to multiple dimensions.

\section{\large Forecasting Methodology}
\label{sec_method_overview}

\begin{figure*}[ht]
	\centering
	\includegraphics[width=\textwidth]{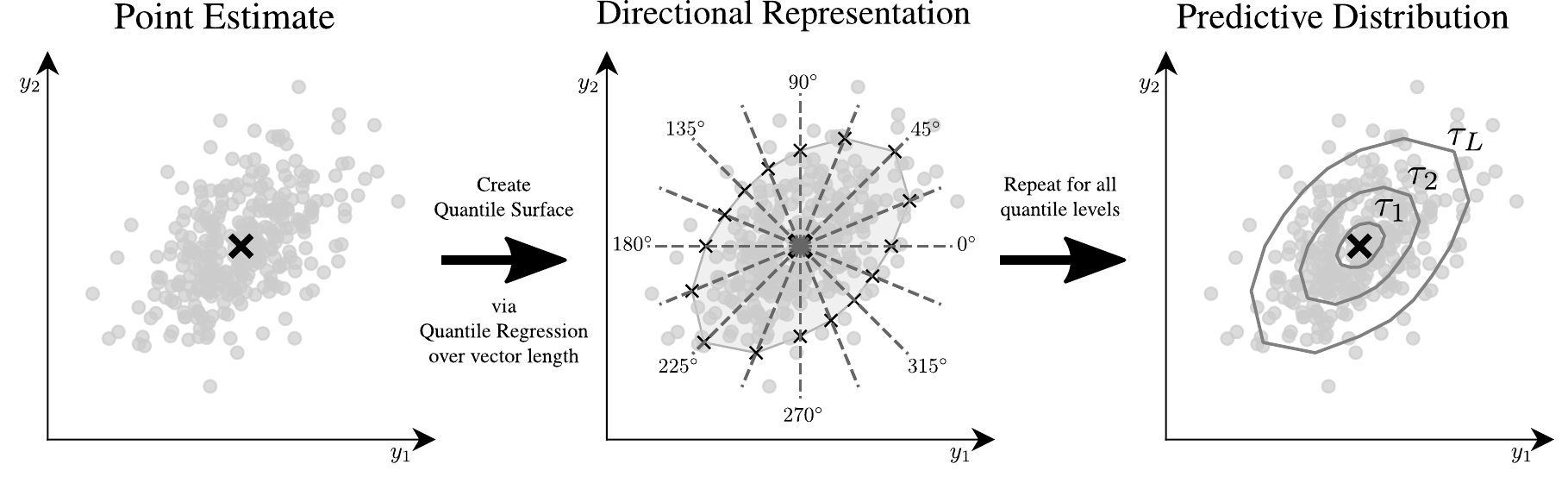}
	\vskip 5mm
	\caption{Illustration of quantile surface approach: First, we perform a deterministic point estimation. Based on this  point estimate, we employ a directional representation of points in the target space, i.e., polar coordinates in the depicted two-dimensional case. The quantile surface for a specific quantile is created by applying quantile regression over the vector length for the various directions. The entire predictive distribution is given by the evaluation of multiple discrete quantile levels $\tau_1, \ldots, \tau_L$.}
	\label{fig:qs_methdology_complete}
\end{figure*}

A natural way of extending quantiles to multiple dimensions is to consider sets incorporating a particular predefined probability mass $\tau$.
Our QS methodology builds upon this intuition. The goal of our QS prediction methodology is to predict enclosing boundaries of a set having a predefined probability mass of $\tau$ while being as compact as possible.
In doing so, we index points in the particular set through the displacement from the estimate of central tendency (e.g., mean, median, or mode). The displacement vector is represented by a direction vector and the corresponding vector length. Using this representation, we can now recover ordering information with respect to the length, i.e., the Euclidean distance, of the displacement vector. We propose to represent multivariate probability distributions by means of discrete quantile surfaces levels $\boldsymbol{\tau} := \left(\tau_1, \ldots, \tau_L \right)$.  We depict a schematic of our approach for the estimation of the QS in Fig.~\ref{fig:qs_methdology_complete}.

To summarize, our approach consists of two stages: (1) a deterministic point estimate and (2) estimates of the discrete quantile surfaces for different quantile levels $\tau_i, i \in (1, \ldots, L)$ with respect to the central tendency estimate. We restrict ourselves to star-shaped, unimodal predictive distributions.
The only assumption for the deterministic point estimate of the first stage is that it has to represent the central tendency of the underlying predictive distribution, e.g., mean or spatial median.

The remainder of this section is structured as follows: First, the methodology for creating deterministic point estimates is detailed. Then we introduce the formal definition of QS as well as the underlying mathematical model. Subsequently, we present the technical realization of QS using neural networks, i.e., quantile surface neural networks (QSNN).



\subsection{Point Estimate} \label{subsec:point_estimate}

In the first stage, we perform a deterministic point estimate using an arbitrary deterministic prediction model, e.g., linear regression, random forest, or ANN. This deterministic point estimate is the starting point of our subsequent probabilistic extensions, i.e., our QS estimates.
Therefore, we require that the deterministic estimate is related to the mean, spatial median~\cite{Cha03}, or the mode of the predictive distribution.
Examples for this are deterministic point estimators that rely on the minimization of a root mean-squared error, a  mean absolute error, or a sum of Euclidean distances.
Note that our approach is not restricted to any specific deterministic prediction model, but instead, our approach enables us to add uncertainty estimates to arbitrary deterministic prediction models. Formally, the $K$ dimensional deterministic point estimate $\hat{\mathbf{y}}_{i}^{(d)}$ of the $i$-th sample is defined as
\begin{align}
\hat{\mathbf{y}}_{i}^{(d)} = f_{\boldsymbol{\theta}}^{(d)} \left(\mathbf{x}_i \right) \in \mathbb{R}^K,
\end{align}
where $f_{\boldsymbol{\theta}}$ represents the deterministic prediction model, $\boldsymbol{\theta}$ its parameters, and $\mathbf{x_i}$ the model input, i.e., features of the $i$-th sample. 


\subsection{Quantile Surface Definition and Model} \label{subsec:quantile_surface}

In the second stage, we aim to predict the uncertainties associated with the deterministic prediction. Here, we rely on QR~\cite{KB78} as a basis for our probabilistic prediction, i.e., we represent the predictive distributions $p \left( \mathbf{y}_i | \mathbf{x}_i \right)$ for a particular sample $i$ by means of distinct quantiles $\tau \in \left]0 , 1\right]$. For a univariate target, each distinct quantile $\tau$ has a single associated value. A natural way of extending this to multivariate targets is to think of sets instead of single values. Our approach is based on the intuition of predicting the boundaries of a set, which comprises an area of probability mass $\tau$. The boundaries or surfaces of these sets are what we refer to as \textit{quantile surfaces} (QS). The overall predictive distribution is now represented by a multitude of these sets. In the further,  we will refer $\tau$-th QS to the QS belonging to quantile $\tau$. However, before we proceed to the concrete realization of QS, we first provide the formal definition of QS and explain the mathematical model behind QS.

Our QS approach uses the following definition of a multivariate quantile function~\cite{SS01, EM92}. 
Let $\mathcal{C}$ be a class of measurable subsets of our target random variable $\mathbf{y}_i$, which is distributed according to $p \left( \mathbf{y}_i | \mathbf{x}_i \right)$.
Let $P_i$ denote the corresponding CDF, and let $\lambda$ be a real-valued function defined on $\mathcal{C}$. The \textit{quantile function} with respect to $\left( P_i, \lambda, \mathcal{C} \right)$ is 
then defined using the infimum $\inf$:
\begin{align}
	U_i \left( \tau \right) = \inf \left\{ \lambda \left( C \right) |\ P_i \left( C \right) \geq \tau, C \in \mathcal{C} \right\}.
	\label{eq:quantile_function}
\end{align}

In our case, $\lambda$ is the Lebesgue measure $\lambda_K$ on $\mathbb{R}^K$. That is, $U_i \left( \tau \right)$ describes the smallest set in $\mathcal{C}$ (where the size is measured by $\lambda$) which has a probability mass of $\tau$. This definition matches our intuition of QS.

\begin{figure}[b]
	\centering
	\includegraphics[width=0.8\linewidth]{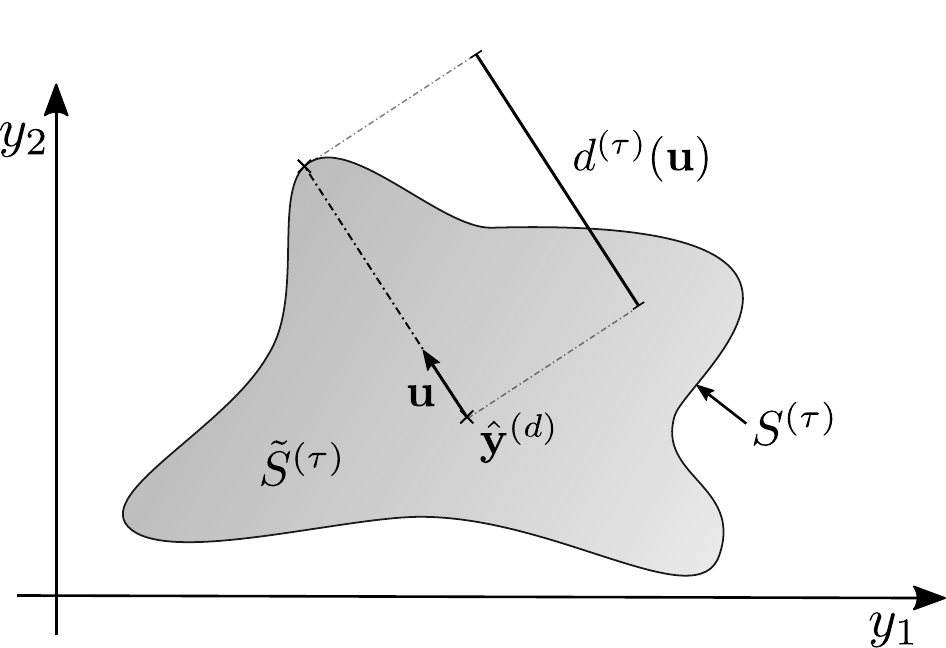}
	\vskip 4mm
	\caption{Schematic of QS $S^{\left(\tau \right)}$ and directional representation of points with respect to a deterministic point forecast describing a star domain set.}
	\label{fig:qs_model}
\end{figure}

In the following, we formalize QS, i.e., the sets and how they are described. A schematic of the explanation described in the following is shown in Fig.~\ref{fig:qs_model}.
From a geometrical perspective, a QS can be viewed as a star domain set in $\mathbb{R}^K$~\cite{Cha77}. 
The origin of the star domain set is a point estimate $\hat{\mathbf{y}}^{(d)}$. The $\tau$-th QS is represented by means of  normalized directional vectors $\mathbf{u} \in \mathbf{Z} :=\left\{ \mathbf{z}\ |\ \mathbf{z} \in \mathbb{R}^K, \left\| \mathbf{z}\right\| = 1 \right\}$ and their length $d^{\left(\tau \right)} \left( \mathbf{u} \right) \in \mathbb{R}$. The length specifies the extent of the \mbox{$\tau$-th} QS in direction $\mathbf{u}$. In the following, this is referred to as directional quantile. 
Formally, we can describe the star domain set $\tilde{S}^{(\tau)}$ as follows:

\begin{align}
\tilde{S}^{(\tau)} := \left\{ \hat{\mathbf{y}}^{(d)} + \gamma \cdot d^{\left(\tau \right)} \left( \mathbf{u} \right) \cdot \mathbf{u}\ |\  \gamma \in \left[0, 1\right], \mathbf{u} \in \mathbf{Z} \right\}.
\label{eq:quantile_surface_set}
\end{align}
The QS $S^{(\tau)}$ is defined as the boundary (or surface) of the set $\tilde{S}^{(\tau)}$. We can obtain the QS by setting $\gamma = 1$.

\begin{figure}[h]
	\centering
	\includegraphics[width=0.8\linewidth]{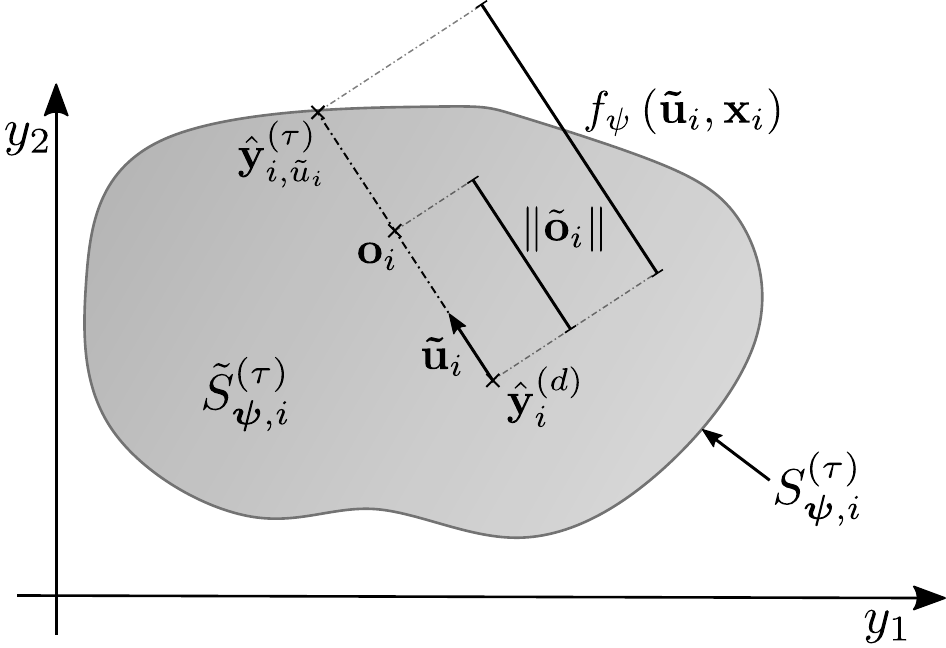}
	\vskip 4mm
	\caption{Schematic of $\tau$-th QS forecast $S_{\boldsymbol{\psi},i}^{\left(\tau \right)}$ and the respective complete set $\tilde{S}_{\boldsymbol{\psi},i}^{\left(\tau \right)}$, whereas $\mathbf{o}_i$ is an observation enclosed within the QS, $\mathbf{\tilde{u}}_i$ the direction, and $\| \tilde{\mathbf{o}}_i \|$ the distance from the deterministic point forecast to the observation.}
	\label{fig:qs_method}
\end{figure}

In the next step, we replace $d^{\left( \tau \right)} \left( \mathbf{u} \right)$ by a function $f_{\boldsymbol{\psi}}^{\left(\tau \right)} \left( \mathbf{u}, \mathbf{x}_i \right)$ which is parameterized by $\boldsymbol{\psi}$. In addition to the orientation, this function also considers the current feature vector $\mathbf{x}_i$ as input. By means of this representation, we can describe conditional star domain sets centered around the point estimate $\hat{\mathbf{y}}_{i}^{(d)}$, whereas, ``conditional'' refers to the dependence on the feature vector $\mathbf{x}_i$. The $\tau$-th QS for the $i$-th sample is now defined as follows:
\begin{align}
S_{\boldsymbol{\psi}, i}^{(\tau)} := \left\{ \hat{\mathbf{y}}_{i}^{(d)} +  f_{\boldsymbol{\psi}}^{\left(\tau \right)} \left( \mathbf{u}, \mathbf{x}_i \right) \cdot \mathbf{u}\ |\  \mathbf{u} \in \mathbf{S} \right\}.
\end{align}
The complete set $\tilde{S}_{\boldsymbol{\psi}, i}^{(\tau)}$ is defined analog to Eq.~\ref{eq:quantile_surface_set}. A schematic of this is depicted in Fig.~\ref{fig:qs_method}.
In the next step, we must choose the model parameters $\boldsymbol{\psi}$ which govern the shape of the QS such that the QS covers at least $\tau$ of the probability mass. In the next section, we show that this can be reduced to a univariate regression task which can be solved through QR. 

An entire QS forecast comprises $L$ distinct quantile levels. Subsequently, the QS forecast $\mathbf{S}_i$ of the $i$-th sample is given by the tuple
\begin{align}
\mathbf{S}_{\boldsymbol{\psi},i}~:=~\left( S_{\boldsymbol{\psi},i}^{\left( \tau_1 \right)}, \ldots, S_{\boldsymbol{\psi},i}^{\left( \tau_L \right)} \right).
\label{eq:qs_forecast_tuple}
\end{align}
For the sake of brevity, when speaking of QS, we omit the parameter vector $\boldsymbol{\psi}$ in the following.

The heart of our QS approach lies in the representation of sets using a representation comprising of a directional component and a one-dimensional vector length.
Note that, in terms of the quantile functions according to Eq.~\ref{eq:quantile_function}, our set of measurable subsets is the set of all possible star domain sets which can be described by our QS. 

\subsection{Technical Realization of Quantile Surfaces and Quantile Surface Neural Networks} \label{subsec:training}

Having formally defined QS, the next question to be answered is how to create, i.e., train and optimize, predictive 
machine learning models that issue sharp and well-calibrated predictive distributions represented using QS.
Based on the previously introduced directional representation of the $\tau$-th QS, we can vary the shape of the predictive 
distribution by adapting the vector length $f_{\boldsymbol{\psi}}^{\left(\tau \right)} \left( \mathbf{u}, \mathbf{x}_i \right)$. 
For each quantile level $\tau$, we need to create a predictive model parameterized by $\boldsymbol{\psi}$, which issues the $\tau$ quantile of the vector length of each direction $\mathbf{u}$ given an input feature vector $\mathbf{x}_i$. Instead of treating the directional vectors separately, we view them as an additional input of our predictive model. The predictive space is, due to the directional representation, no longer multivariate but univariate now. Hence, we can apply univariate QR~\cite{KB78} to obtain quantile estimates for the vector length for each direction, i.e., directional quantile estimates. By using a discrete set of directions, we can obtain a sample-based representation of the $\tau$-th QS for a particular feature $\mathbf{x}_i$.

\begin{figure}[t]
	\centering
	\includegraphics[width=\linewidth, clip, trim = 0 0 0 0]{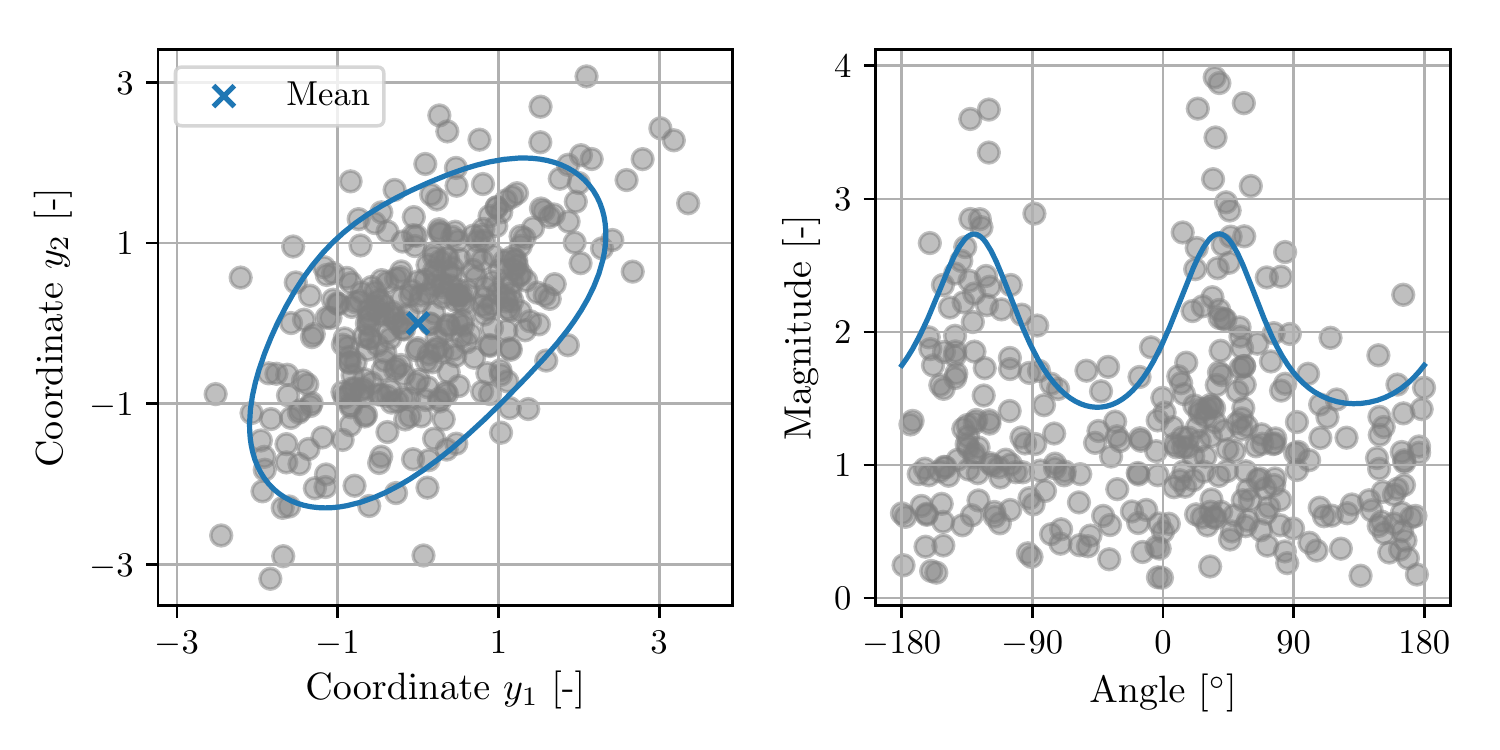}
	\vskip 2mm
	\caption{Visualization of QS for unconditional distributions. On the left: the sample-based representation of the $0.90$-th QS in two-dimensions, on the right: directional representation of the corresponding QS. The angle is measured with respect to the first coordinate axis $y_1$.}
	\label{fig:directional_indexing}
\end{figure}

Consider finding the $0.90$ QS of an unconditional two-dimensional Gaussian distribution with zero mean as depicted in Fig~\ref{fig:directional_indexing}.
As depicted on the right-hand side, the directional indexing allows imagining the QS as a one-dimensional function depending on the angle.
In this case, the fitted function corresponds to the conditional 90\% quantile. 

QR is based on the quantile loss (also referred to as pinball loss or absolute tiled value function). It is an asymmetric weighting of positive and negative errors using a tilted form of the absolute value error function.
The quantile loss is defined for a particular quantile value $\tau$.
Then, given a univariate observation $o$ and the predicted value of the $\tau$-th quantile $\hat{y}$, the quantile loss is defined by
\begin{align}
s_{\tau} \left( o, \hat{y} \right)  = \left\{\begin{array}{cl} \tau | \hat{y} - y |, & \left( \hat{y} - o \right) \geq 0 \\
\left(1 - \tau \right) | \hat{y} - o |, & \text{else}\end{array}\right..
\end{align}
The optimization of this loss function adjusts the model parameters such that the issued predictions correspond to the $\tau$-th quantiles of the underlying distribution. 
As shown in~\cite{Gen19}, the quantile loss reaches its minimum at the true $\tau$ quantile of the underlying distribution.
The complete predictive distribution (represented using discrete quantiles) can now be obtained by training $L$ distinct models, i.e., one for each quantile level. Originally, QR was tailored and designed for linear models~\cite{KB78}. However, using the pinball loss, QR can also be extended to non-linear models such as ANN (here referred to as QRNN) or gradient boosted regression~\cite{Fri01}.

\begin{figure}
	\centering
	\includegraphics[width=0.8\linewidth, clip, trim=100 650 100 50]{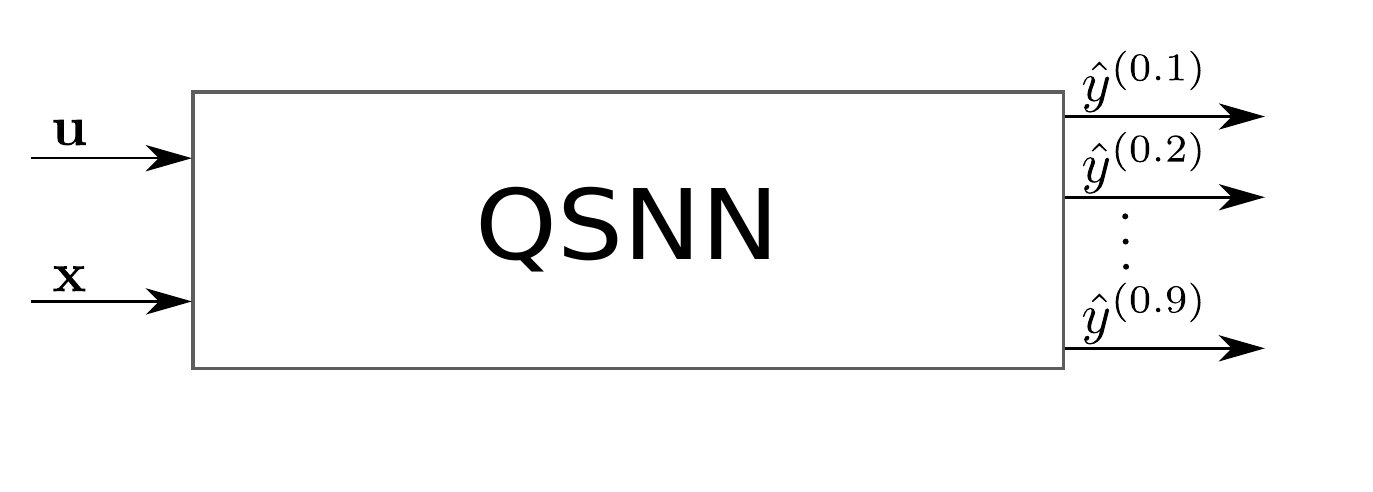}
	\caption{Schematic of QSNN for prediction of multiple quantile levels at the same time. We obtain a sample-based representation of the entire QS forecast via the repeated evaluation of the QSNN using a discrete set of directions $\mathbf{u}$.}
	\label{fig:qrnn}
\end{figure}

Let  $\left( \mathbf{x}_i, \mathbf{o}_i \right)$ be the $i$-th training sample consisting of a feature vector $\mathbf{x}_i$ and the corresponding observation $\mathbf{o}_i$. The observation is our target variable for which we aim to predict.
 We represent an observation utilizing a directional vector, i.e., a vector with the unit norm, and associated vector length, with respect to the deterministic central tendency estimate. In doing so, we subtract the deterministic point estimate from our observation:
\begin{align}
\tilde{\mathbf{o}}_i = \mathbf{o}_i - \hat{\mathbf{y}}^{\left(d \right)}_i.
\label{eq:forecast-adjusted_predictions}
\end{align}
In the following, this transformed observation is referred to as forecast-adjusted observation. A visualization of this is depicted in Fig.~\ref{fig:qs_method}. To train the QR model such that it estimates the vector length's quantiles, we need to recover the directional information corresponding to the particular observation $\tilde{\mathbf{o}}_i$. This is done by 
\begin{align}
\mathbf{\tilde{u}}_i = \frac{\tilde{\mathbf{o}}_i}{\| \tilde{\mathbf{o}}_i \|}.
\end{align}
Next, we have it all in place to define the objective function to train models $f_{\boldsymbol{\psi}} \left( \mathbf{\tilde{u}}_i, \mathbf{x}_i  \right)$ to estimate the QS. The objective function used in our QS approach is given by the quantile loss applied to the vector length of the forecast-adjusted observation and the predicted vector length's quantile. The resulting objective function for all $N \in \mathbb{N}$ samples is given by 
\begin{align}
E \left( \boldsymbol{\psi} \right)  = \sum\limits_{i = 1}^{N} s_{\tau} \left( \| \tilde{\mathbf{o}}_i \|,  f_{\boldsymbol{\psi}} \left( \mathbf{\tilde{u}}_i, \mathbf{x}_i  \right) \right).
\end{align}
In general, we could use this objective function to optimize various models, such as linear QR model or QRNN, to estimate the directional quantile length and, therefore, predict the shape of the QS. However, we focus our investigation on QRNN, since neural networks are a flexible and widely adopted methodology which is easy to customize and allows for modeling arbitrary functions. 
For example, we could enforce the positivity of predicted directional quantiles by means of a rectified linear or exponential output unit.
We could train $L$ distinct models to obtain QS predictions for each level $\tau$. 
Nevertheless, having many individually trained models often leads to quantile crossing, i.e., a lower quantile protrudes a higher quantile~\cite{HAS+17}. Although we cannot fully alleviate the effect of quantile crossing, we can reduce it. Instead of training models for each distinct quantile $\tau$, 
we train one single neural network with $L$ outputs, one for each quantile level. We term this neural network quantile surface neural network (QSNN). A schematic of the QSNN is depicted in Fig.~\ref{fig:qrnn}. We train the model with $L$ different loss functions.
This approach realizes a weight-sharing which regularizes the model and henceforth increases the generalization ability to unseen data.  Besides, as post-processing we sort the estimated quantiles in ascending order. This simple heuristic avoids quantile crossing. More advanced techniques to avoid quantile crossing, with stronger theoretical foundations, e.g., the incorporation of the distance between neighboring quantiles in the loss function, can be found in~\cite{HAS+17}.

 

%
%

%% file: sections/evaluation_method.tex
\section{Evaluation Methodology} \label{sec:evaluation_method}
Central aspects of probabilistic forecasts are \textit{reliability} and \textit{sharpness}. Reliability refers to whether the issued predictive distribution is correct, i.e., whether the estimated frequency of the predictive distribution matches the observed frequency. 
A reliable forecast is also referred to as a well-calibrated forecast.
Sharpness gives insights into the quality of the forecasts, i.e., it measures the narrowness of the predicted probability distribution. The ultimate goal of training models for probabilistic forecasts is to maximize the sharpness while retaining reliability. These are also referred to as scoring rules, which together assess the \textit{skill} of a probabilistic forecasting model.
To be able to assess whether the trained models issue sharp and well-calibrated predictive distributions using QS, we need to define the objective to measure both simultaneously. This is also referred to as the skill of the forecasting model. We can assess the skill by loss functions, such as the quantile score~\cite{KB78}. 
The quantile score evaluates a single quantile $\tau$, \textit{not} an entire predictive distribution.
However, in many cases, especially in model selection, the form of the complete predictive distribution is of interest.
Here, the continuously ranked probability score (CRPS) can be used to assess the skill, i.e., reliability and sharpness, with respect to the complete predictive distribution.


\subsection{Assessing the Reliability of QS}
The reliability of a probabilistic quantile forecast describes whether the nominal quantile value matches the observed relative frequency of occurrences within the boundaries of the predicted quantile. This must account for all quantiles. In terms of QS, this means the following: Consider, for example, a forecast of the 0.8 QS, the observed frequency of an observation being within the QS should then ideally be 80\%.

Assume we are given a set of observations, which we denote as $\mathcal{O}= \left( \mathbf{o}_1, \ldots, \mathbf{o}_N \right)$, and a corresponding set of QS forecasts $\mathcal{QS} = \left( \mathbf{S}_1, \ldots, \mathbf{S}_N \right)$, whereas each QS forecast comprises $L$ distinct quantile levels (cf. Eq.~\ref{eq:qs_forecast_tuple}). 
A QS is, by definition, a star domain set with the deterministic point estimate being the origin. Every other point (i.e., observation) in the QS can be connected via a line with the origin with no line segment leaving the QS. To check whether an observation is in the QS, we trace the line between the origin and the observation. Then we compare the predicted length of the quantile in the direction of the observation with the distance between the origin and the observation along that line. If this distance does not exceed the corresponding predicted length, then the observation is contained within the QS. Hence, the frequency of observations falling within a predicted QS can now be determined using a sum of values of the Heavyside functions $H$:
\begin{align}
	v^{\left( \tau \right)} = \frac{1}{N} \sum\limits_{i = 1}^{N} H \left( f_{\boldsymbol{\psi}} \left( \mathbf{\tilde{u}}_i, \mathbf{x}_i \right) - \|\tilde{\mathbf{o}}_i\| \right),
\end{align}
with $f_{\boldsymbol{\psi}} \left( \mathbf{\tilde{u}}_i, \mathbf{x}_i \right)$ being the estimated $\tau$-th directional quantile of the $i$-th sample in direction $\tilde{\mathbf{u}}_i$ and $\|\tilde{\mathbf{o}}_i\|$ being the distance from the deterministic point estimate to the corresponding observation. Hence, evaluating the reliability of multivariate QS forecasts can be reduced to the evaluation of univariate QR. 
We can then evaluate the reliability of the overall predictive distribution visually using a quantile-quantile plot (Q-Q-plot) by plotting the observed frequency against the expected frequency~\cite{mbGK14}. 


\subsection{Assessing the Sharpness of QS}

The second key property of probabilistic forecasts is their sharpness. Sharpness refers to the concentration of 
the distribution and, therefore, is a property exclusive to the forecasts, i.e., it can be evaluated without consideration of the observations~\cite{mbGK14}. 
Thus, it refers to the quality of the forecast, i.e., the ability of the model to issue narrow predictive distributions. 

For univariate predictive distributions, the sharpness can be evaluated considering the average size of the intervals spanned for the dedicated nominal coverage probability levels~\cite{Gen19}. In terms of multivariate probabilistic forecasts and QS, the sharpness refers to the area or (hyper-)volume covered by the predictive distribution. We must now consider the size of the (hyper-)volume covered by the predictive distribution for a particular nominal level.

Let $\lambda$ be the Lebesgue measure, then $\lambda \left( S_i^{\tau} \right)$ describes 
the volume of the $\tau$-th QS of the $i$-th sample.
We can derive a measures of sharpness $\kappa_{QR}^{\left( \alpha \right)}$ for a particular nominal coverage probability $\left( 1 - \alpha\right)$ with $\alpha \in \left[0, 1 \right]$:
\begin{align}
	\kappa_{QS}^{\left( \alpha \right)} = \frac{1}{N} \sum\limits_{n=1}^N \lambda \left( S_i^{\tau_{\hat{v}}} \right).
	\label{eq:sharpness_kappa}
\end{align}
In Eq.~\ref{eq:sharpness_kappa}, we set $\tau_{\hat{v}} = 1 - \alpha$. This formula allows us to compute a measure of sharpness for QS which is directly related to the average area or volume covered by the $\tau_{\hat{v}}$-QS forecasts. If we use different nominal coverage probabilities (i.e., different $\alpha$), we can compute an average sharpness measure~\cite{Gen19}. In this article, we use these measures to evaluate the sharpness visually~\cite{mbGK14, JFK08}.
Still, the question is how to compute the Lebesgue measure, i.e., how to determine the (hyper-)volume of the issued QS.
At best of our knowledge, there is no straightforward, exact, and computationally efficient way to compute the volume spanned by a general $K$-dimensional QS. In this case, we refer to approximations, such as Monte-Carlo sampling-based approximations~\cite{KU98}. However, for two dimensions and, thus, for many applications, there is an efficient algorithm. It is based on the interpretation of the QS as a planar polygon, for which we can apply the surveyor's area formula~\cite{Bra86} to evaluate the QS. In the applications examples considered in this article, we limit ourselves to two-dimensional problems.

\subsection{Assessing the Skill of QS -- Directional CRPS}
The skill quantifies the error of a probabilistic forecasting technique on a data set.
It incorporates an error, which can be attributed to both reliability and sharpness.
Among other scores, such as the quantile and ignorance score~\cite{Gen19}, the CRPS is one of the most widely applied scores for skill assessment of probabilistic forecasts. The CRPS is an evaluation measure for probabilistic prediction models with a one-dimensional target quantity. It can be considered as a generalization of the mean absolute error (MAE) for probabilistic predictions. In contrast to the MAE, for probabilistic forecasts it is not sufficient to evaluate the pure difference between the prediction and the observation. Instead, we must assess the deviation between the predictive distribution and the observations. The CRPS achieves this by modeling the CDF of the observation in the form of a Heaviside step function $H$ and evaluating the difference between the observed $H(y -o)$ and the predicted CDF $\hat{P}(y)$. It is given by
\begin{align}
\text{CRPS}\,(\hat{P}, o) = \int_{-\infty}^{\infty} \left(\hat{P}(y) - H \left(y - o \right) \right)^2 \mathrm{d}y.
\label{eq:crps}
\end{align}
The multivariate extension of the CRPS is the energy score~\cite{GR07}. 
Yet, the computation for QS is tedious. Furthermore, there is currently no known possibility to evaluate the energy score for QS. Therefore, we propose an alternative variant of the CRPS for the evaluation of multivariate quantiles. We term this score, which can be computed extremely efficiently, \textit{directional CRPS}. The predictive distribution is evaluated with respect to the directions of samples. Instead of evaluating the complete multivariate predictive distribution, we instead consider the univariate CRPS for the predicted CDF of the vector length.

Therefore, let $\tilde{\mathbf{u}}$ denote direction and $\|\tilde{\mathbf{o}}\|$ the magnitude of the directional representation of the observation $\mathbf{o}$ with respect to the deterministic central tendency estimate  $\mathbf{\hat{y}}^{\left(d\right)}$.  
$\hat{P}_{\tilde{\mathbf{u}}} \left( y \right)$ denotes the predicted CDF of the vector length with respect to direction $\tilde{\mathbf{u}}$ (i.e., a particular direction of the QS in $\mathbf{\hat{y}}^{\left(p \right)}$).
Using this, the directional CRPS is defined as
\begin{align}
	 \text{CRPS}_{\text{\,DIR}} (\hat{P}_{\tilde{\mathbf{u}}}, \|\tilde{\mathbf{o}}\|) = \int_{-\infty}^{\infty} \left( \hat{P}_{\tilde{\mathbf{u}}} \left( y \right) 
	- H \left(y - \|\tilde{\mathbf{o}}\| \right) \right)^2 \mathrm{d}y,
	\label{eq:directional_crps}
\end{align}
where $H$ denotes the Heavyside step function.

Since the score only evaluates the issued predictive distribution for a particular direction, one could argue that the score neglects parts (i.e., directions) of the predictive distribution where there are no observations for evaluation. The evaluation of forecasts with no reference, i.e., no observations, is always difficult. In this case, we can only evaluate the aspects relating to the predictive distribution itself, e.g., the sharpness. The directional CRPS does not cover those parts of the predictive distribution in the score. However, as we are assuming that our point of origin is a central tendency estimate, we expect
that all directions (given by the observations) occur approximately with equal frequency, i.e., it is unlikely that we are neglecting vast parts in the evaluation of the predictive distribution where there are no observations. 

So far, we have not discussed how to compute the directional CRPS for QS. As we are only considering the direction and the length, we can omit all quantile estimates for other directions in the QS except the direction of the respective observation. For the respective direction, we evaluate the directional CRPS as depicted in Eq.\ref{eq:directional_crps}.
To evaluate the integral, we can obtain the univariate CDF of the set of $L$ distinct quantile forecasts~\cite{Gen19}.
Subsequently, we efficiently approximate the directional CRPS using the methodology described in~\cite{Her20}.

The directional CRPS can be computed for arbitrary probabilistic forecasts for which we can assess the predictive distribution over the vector length with respect to the direction. This includes, in particular, the Gaussian distribution, in which the directional CRPS can be assessed by means of the Mahalanobis distance~\cite{Bis06, Bre15} of the observation with respect to the mean of the distribution. 

The univariate CRPS is related to the MAE. Hence, by construction, the directional CRPS is also related to the directional MAE, i.e., the MAE calculated for the vector length of the directional representation. In our case, this corresponds to the Euclidean norm between the deterministic central tendency estimate and the observation. We can use this relationship between the directional CRPS and the directional MAE, i.e., the Euclidean norm, for a direct comparison of multivariate deterministic and multivariate probabilistic forecasting methodologies.

To evaluate $N  \in \mathbb{N}$ forecasts and observations, we use the average directional CRPS.
It is defined as 
\begin{align}
\overline{\text{CRPS}}_{\text{\,DIR}} = \frac{1}{N}\sum_{i = 1}^{N}  \text{CRPS}_{\text{\,DIR}} (\hat{P}_{i, \tilde{\mathbf{u}_i}},  \|\tilde{\mathbf{o}}_i\|),
\end{align}
where $\hat{P}_{i, \tilde{\mathbf{u}_i}}$ denotes the predicted CDF (e.g., obtained from a QS forecast) of the vector length of the $i$-th sample in direction $\mathbf{u}_i$. As before, $\|\tilde{\mathbf{o}}_i \|$ denotes the vector length from of the $i$-th observation to the deterministic estimate of central tendency.

\subsection{Multivariate Gaussians as Baseline} \label{subsec:multivariate_gaussian_baseline}
As baseline, we assume that the target variable is distributed according to a multivariate Gaussian distribution~\cite{Bis06} 
with mean  $\boldsymbol{\mu} \in \mathbb{R}^K$ and covariance $\boldsymbol{\Sigma} \in \mathbb{R}^{K \times K}$.
We distinguish two cases: First, an unconditional Gaussian, which in this case means that the covariance $\boldsymbol{\Sigma}$ remains fixed and is independent of a feature input (i.e., homoscedasticity). Second, we also consider the conditional case in which besides the mean $\boldsymbol{\mu}$ also the covariance $\boldsymbol{\Sigma}$ depends on the current input $\mathbf{x}$ (i.e., heteroscedasticity). To model the dependence of the mean and the covariance on the input, we use a neural network. This is a neural network that takes a feature vector as input and outputs the parameters of a multivariate Gaussian distribution~\cite{Bis06}.

To evaluate reliability, sharpness, and skill (i.e., directional CRPS) of the baseline model, we need to 
assess the quantiles of the multivariate Gaussian.
Therefore, we use the squared Mahalanobis distance and the Chi-square ($\chi^2$) distribution. 
As shown in~\cite{Bre15}, the squared Mahalanobis distances based on a Gaussian distribution are $\chi^2$ distributed.
Subsequently, we can use the inverse CDF of the $\chi^2$ to assess the quantiles.

The Mahalanobis distance between the mean of the predictive distribution and 
an observation $\mathbf{o}$ is given by

\begin{align}
d_{\text{M}} \left(\boldsymbol{o} , \boldsymbol{\mu} \right) = \sqrt{\left(\mathbf{o} - \boldsymbol{\mu}\right)^\text{T} \boldsymbol{\Sigma^{-1}} \left(\mathbf{o} - \boldsymbol{\mu}\right) }
\end{align}
The univariate normal distribution covers for $\mu \pm k \sigma, k=1, 2,3$ standard 
deviations $\sigma$ (a Mahalanobis distance of one, two, and three) $68\%, 95\%$, and $99.7\% $ of 
the probability mass, respectively. 

%
Being able to assess the quantiles, we can now compute the directional CRPS for our baseline model, i.e., the multivariate unconditional Gaussian with fixed covariance. 
We measure the skill of the QS forecasting methodology by evaluating the average directional CRPS with respect to 
this baseline. Therefore, we evaluate the skill score, which is generally defined

\begin{align}
\text{Skill} = 1 - \frac{\overline{\mathrm{S}}_{\text{eval}}}{\overline{\mathrm{S}}_{\text{base}}},
\end{align}
where $\overline{\mathrm{S}}_{\text{eval}}$ is the average directional CRPS of the model considered for evaluation (here the QS methodology), and $\overline{\mathrm{S}}_{\text{base}}$ the average directional CRPS of the baseline model.

%

%% file: sections/results_outline.tex
\section{Experimental results}
\label{sec_ResultsOutline}
We evaluate our novel probabilistic forecasting methodology on synthetic as well as on two real-world data sets.
We compare our QS methodology to a baseline method comprising a Gaussian model. The considered synthetic data set constitutes different unconditionally and conditionally distributed targets. 
In doing so, we aim to evaluate the general functionality and performance of our forecasting methodology.
The real-world data sets originate from two different application domains: First, wind power forecasting with a particular focus on modeling the relationships between different wind parks using QS, Second, multivariate trajectory forecasting of cyclists 
enabling active safety applications and automated driving in urban areas.

\subsection{Synthetic Data}
\input{sections/results_synthetic_data}

\subsection{Wind Power Forecasting}
\input{sections/results_wind_power_forecasting}

\subsection{Cyclist Trajectory Forecasting}
\input{sections/results_cyclist_trajectory_forecasting}

%% file: sections/results_synthetic_data.tex
First, we evaluate the properties of our QS approach with synthetic data. 
The considered synthetic data sets include unconditionally and conditionally Gaussian distributed as well as non-Gaussian distributed targets. As baseline, we model the predictive distribution using an unconditional as well as a conditional multivariate Gaussian distribution.
For the synthetic data sets with Gaussian distributed targets, the baseline, an (un)conditional Gaussian distribution, is the best possible model, whereas, for the non-Gaussian distributed targets, the (un)conditional Gaussian distribution is expected to perform worse. The parameters of the baseline model are estimated based on the data using maximum likelihood estimates. For those datasets, for which the ground truth distribution is Gaussian and known, we directly consider the parameter of the data generating process. Both, our QS approach and the baseline share the same deterministic central tendency estimate.


\begin{figure*}[t]
	\centering
	\vskip -4mm
	\begin{subfigure}[t]{0.325\linewidth}
		\centering
		\includegraphics[width=\textwidth]{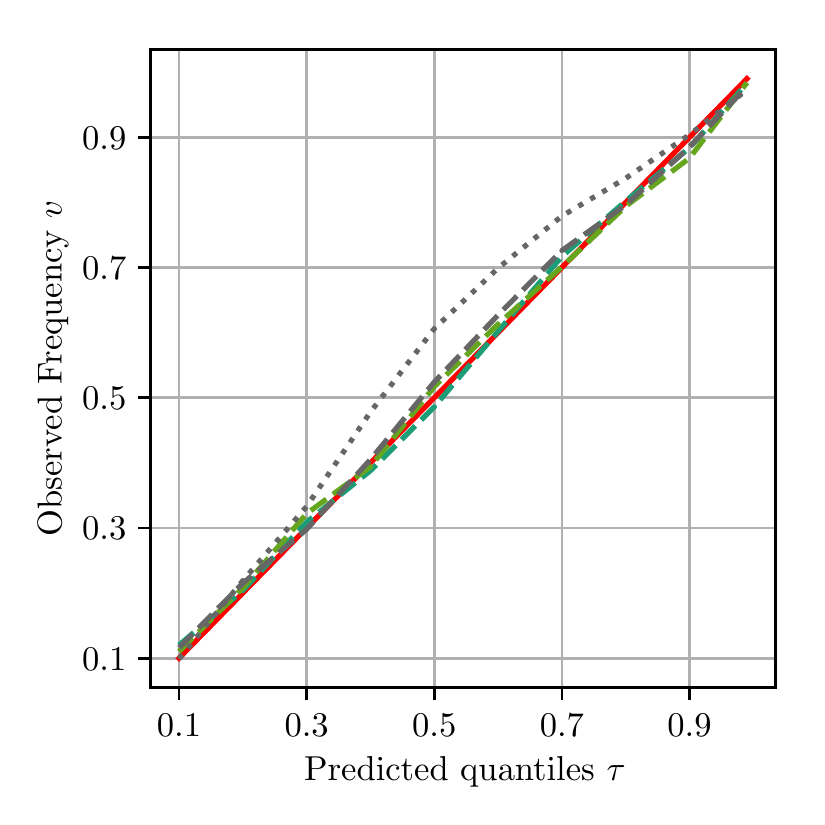}
		\caption{Q-Q-plot evaluating calibration and reliability.}
		\label{fig:q-q-plot_synthetic}
	\end{subfigure}
	\begin{subfigure}[t]{0.325\linewidth}
		\centering
		\includegraphics[width=\textwidth]{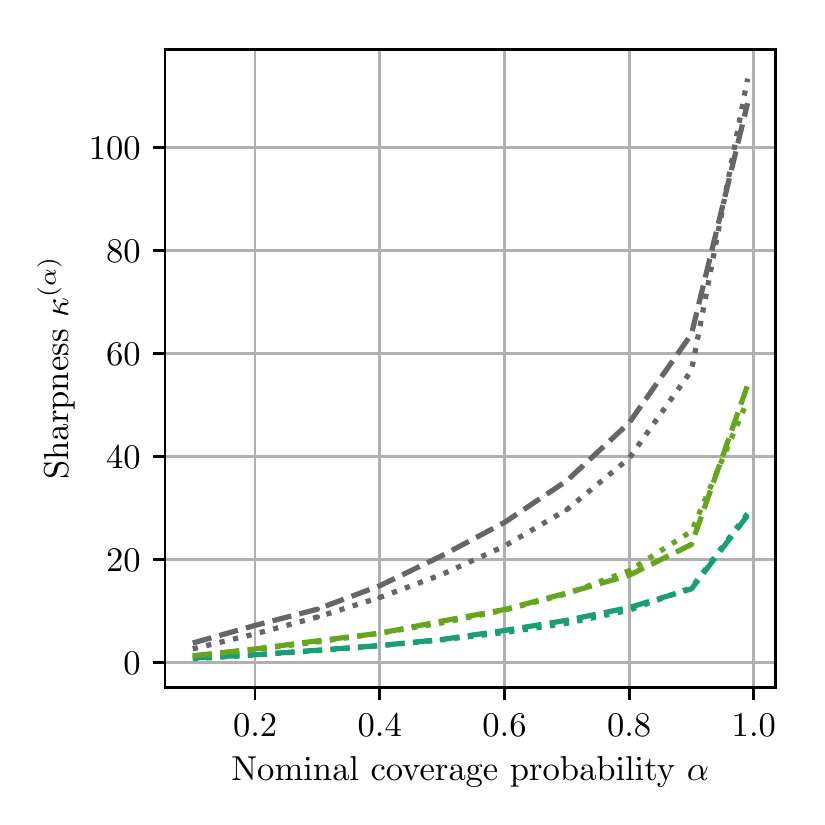}
		\caption{Sharpness diagram: Size of the covered area.}
		\label{fig:sharpness_mvn}
	\end{subfigure}
	\begin{subfigure}[t]{0.325\linewidth}
		\centering
		\includegraphics[width=\textwidth]{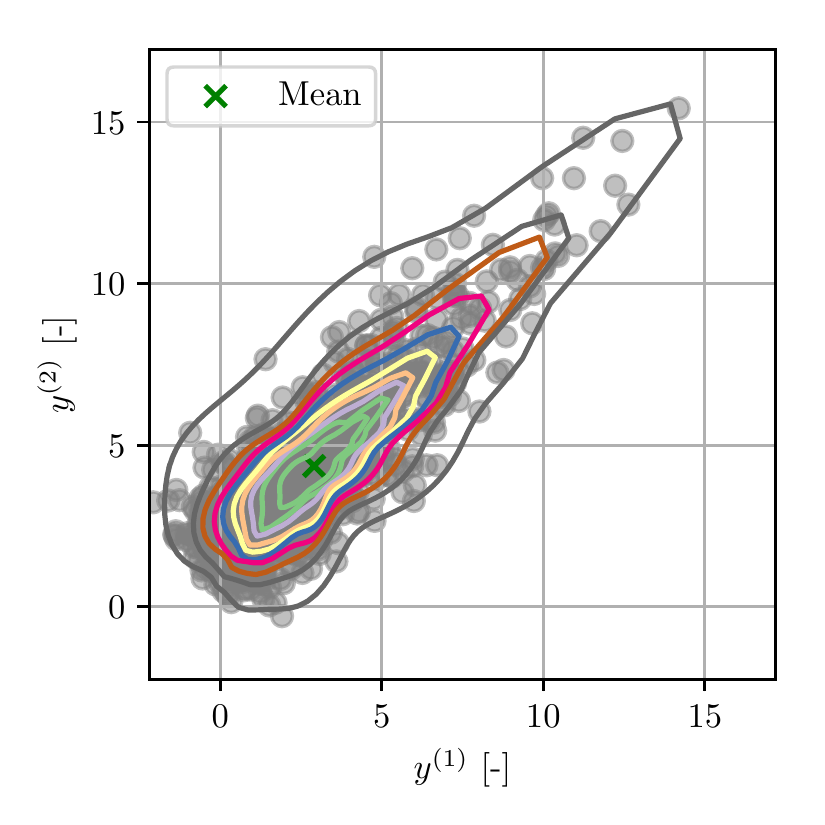}
		\caption{Estimated QS for two-dimensional skewed distribution for different quantile levels.}
		\label{fig:skewed_dist}
	\end{subfigure}

	\vskip 5mm
	\caption{
		Results of the QS forecasting methodology on three different synthetic datasets (Green: MGD, Gray: SMD, and Cyan: CMGD).
		The dashed lines correspond to the QS approach on the respective synethetic data set, while the dotted lines are the baseline models.
		In (c), samples drawn from a two-dimensional skewed distribution (SMD) and QS are shown for different quantile levels $\tau \in \left\{0.1, \ldots, 0.9, 0.99\right\}$ estimated by a QRNN.}
	\label{fig:synthetic_eval}
	\vskip -2mm
\end{figure*}

For our QS approach, we considered a QSNN comprising a single hidden layer with $10$ hidden neurons. As activation function, we apply a hyperbolic tangent. The output neurons are linear units. 
We trained the network for $50000$ epochs using stochastic gradient descent (i.e., the Adam optimizer~\cite{KB15}), with an initial learning rate set to $0.1$. Additionally, we apply L2-regularization with the regularization parameter $\lambda$ set to $0.3$. 

For each of the three synthetic evaluation examples presented in the following, we consider $1000$ samples for training and another $1000$ for testing.

\paragraph{Multivariate Gaussian Distribution (MGD)}
In this example, we demonstrate the general ability of QS to model a two-dimensional Gaussian distribution with 
mean $\boldsymbol{\mu} = \mathbf{0}$. We set the variance of the first axis $\sigma^{\left( 1 \right)} = \sqrt{0.5}$ and the second axis $\sigma^{\left( 2 \right)} = \sqrt{2}$. The covariance between the first and second dimension is set to zero.
Since we know the data generating process, we set it as our baseline. 
As we can see in the Q-Q-plot in Fig.~\ref{fig:q-q-plot_synthetic}, the issued predictions of the QS approach are well-calibrated, i.e., close to the theoretical quantiles. Since the baseline is the data generating process, it is omitted in the Q-Q-plot. Moreover, as can be seen in Fig.~\ref{fig:sharpness_mvn}, the sharpness of the forecasts is congruent with the baseline.

\paragraph{Skewed Multivariate Distribution (SMD)}
In this synthetic evaluation case, we demonstrate the ability of our QS methodology to model multivariate skewed unconditional probability distributions. Therefore, we consider the following construction of a two-dimensional skewed distribution. The first dimension  $y^{\left(1\right)}$ is sampled from an univariate Gaussian with mean $\mu^{\left( 1 \right)} = 1$ and $\sigma^{\left(1 \right)} = 3$, while the second dimension $y^{\left(1\right)}$ is sampled from an exponential distribution with the scale parameter $\lambda$ set to $4$.
Subsequently, we interpret both $y^{\left(1\right)}$ and $y^{\left(2\right)}$ as being part of a two-dimensional random vector. This vector is rotated by 
$45^{\circ}$ to obtain the final sample. As can be seen in Fig.~\ref{fig:skewed_dist}, the resulting distribution is heavily skewed.
We consider the maximum likelihood estimate for the mean as the deterministic point prediction of our QS methodology.
Our QS prediction methodology is able to capture this skewness, i.e., the QSNN adapts to the data. For smaller $\tau$, the 
issued QSs roughly correspond to Gaussians (i.e., they are elliptical), while the QS corresponding to large quantile levels capture the skewness.
As depicted in the Q-Q-plot in Fig.~\ref{fig:q-q-plot_synthetic}, the forecasts issued by our QS forecasting methodology are well-calibrated.
The baseline is obtained by fitting a multivariate Gaussian distribution using a maximum likelihood approach for the mean and covariance. 
In contrast to the QS approach, the skewness is not captured by the multivariate Gaussian. This can be seen in the Q-Q-plot (gray dotted line) depicted in Fig.~\ref{fig:q-q-plot_synthetic}. Here, the baseline issues underconfident predictions. The sharpnesses of both approaches are comparable.

\paragraph{Conditional Multivariate Gaussian Distribution (CMGD)}
In this example, we show the capability of the prediction methodology to issue well calibrated predictive distributions 
for conditional distributions. Therefore, we consider the following conditional two-dimensional Gaussian distribution, where the covariance matrix depends on the one-dimensional binary input feature $x_{cond}$. If the feature is zero, then the variances are set to $\sigma^{\left(1 \right)} = \sqrt{0.5}$ and $\sigma^{\left(2 \right)} = \sqrt{7.5}$. If the feature is one, then the variances are set 
to $\sigma^{\left(1 \right)} = \sqrt{5.0}$ and $\sigma^{\left(2 \right)} = \sqrt{0.5}$. 
In both cases, the mean is set to zero and no covariances are considered. We sample the same number of samples for both conditions, i.e., $500$.
As baseline, we consider the data generating process. As can be seen in Fig.~\ref{fig:q-q-plot_synthetic}, the predictions issued by our QS approach are well-calibrated. The sharpness coincides with the baseline (cf. Fig.~\ref{fig:sharpness_mvn}). Hence, our approach is able to issue well-calibrated conditional predictive distributions.
%
%



%

%% file: sections/results_wind_power_forecasting.tex
\begin{figure*}[ht!]
	\centering
	\begin{subfigure}[t]{0.32\linewidth}
		\centering
		\includegraphics[width=0.95\columnwidth]{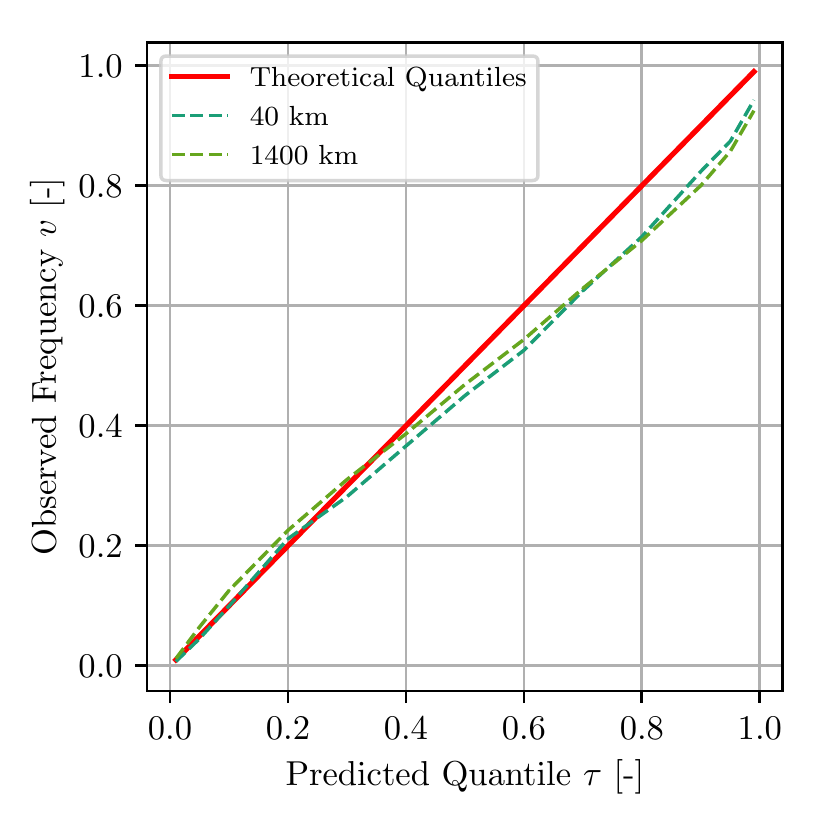}
		\caption{Q-Q-Plot for parks with a distance of $40$\si\km~and $1400$\si\km.}
		\label{fig_wind_qq}
	\end{subfigure}
	\begin{subfigure}[t]{0.32\linewidth}
		\centering
		\includegraphics[width=0.95\columnwidth]{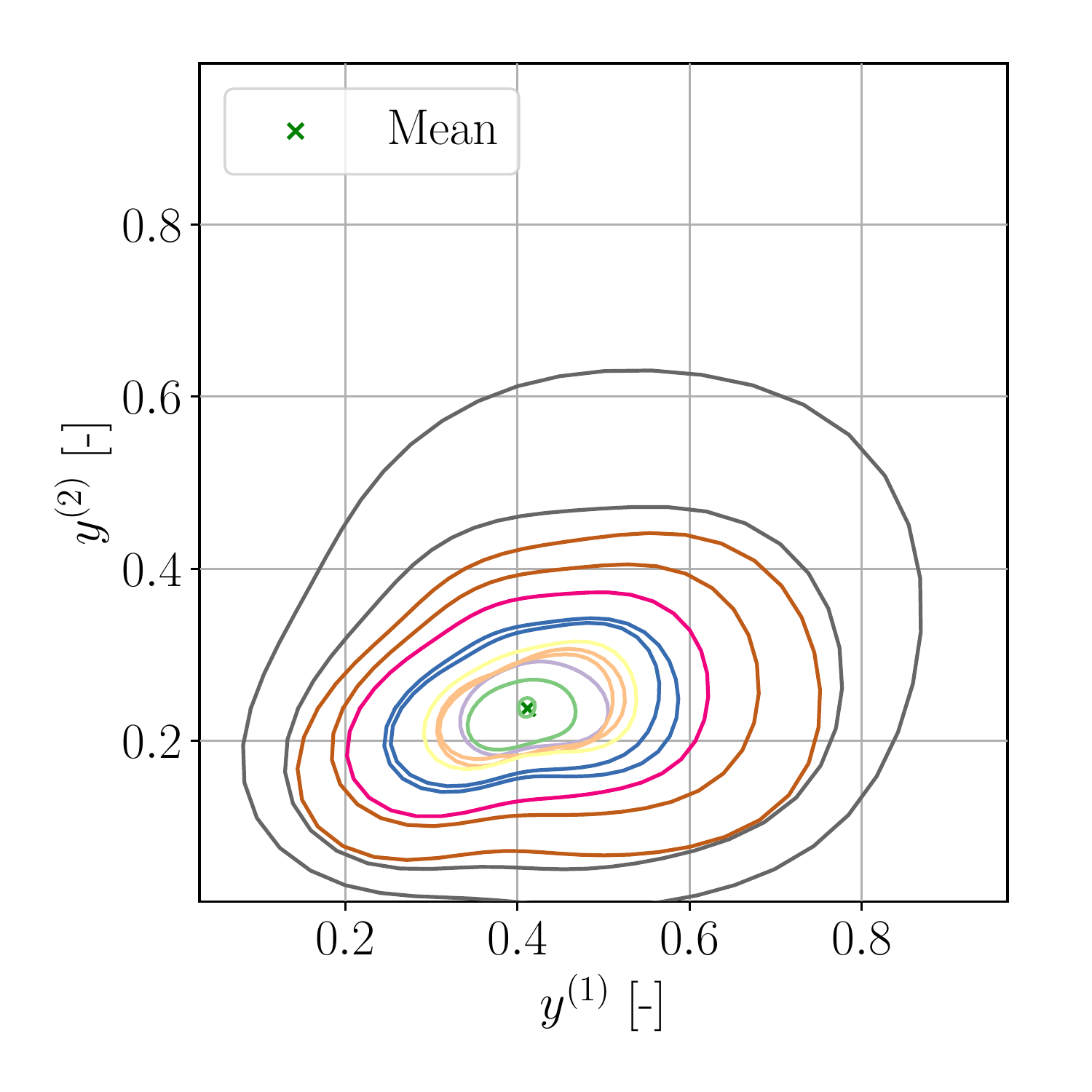}
		\caption{Sample of conditional forecast of farms with $40$\si\km~distance.}
		\label{fig_40km_sample}
	\end{subfigure}
	\begin{subfigure}[t]{0.32\linewidth}
		\centering
		\includegraphics[width=0.95\columnwidth]{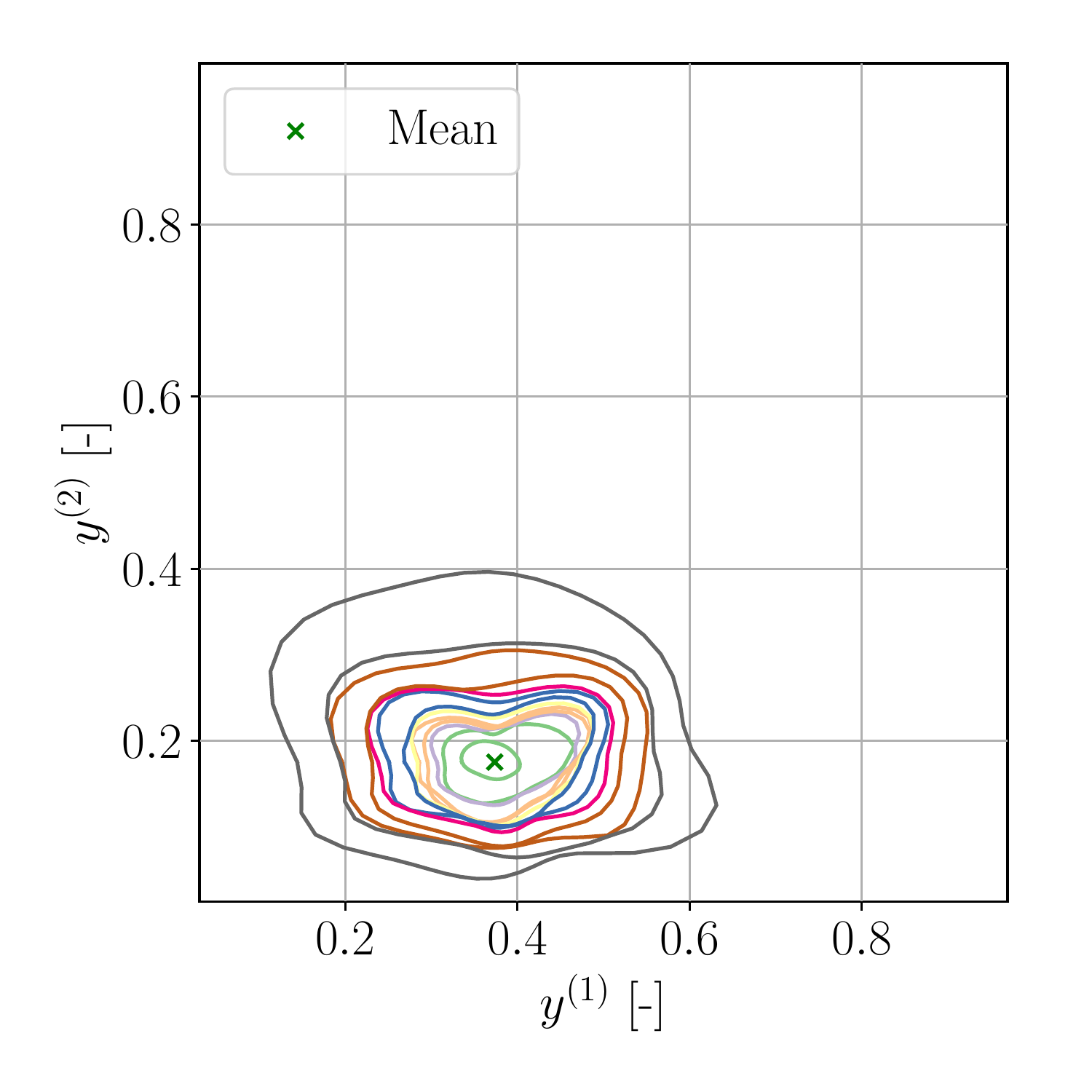}
		\caption{Sample of conditional forecast of farms with $1400$\si\km~distance.}
		\label{fig_1400km_sample}
	\end{subfigure}
	\vskip 5mm
	\caption{
		Q-Q-Plot and QS samples for conditional forecasts issued by the trained QRNN for wind power forecasting.
	}
	\label{fig_wind_results_samples}
\end{figure*}

Recently, probabilistic forecasts can be seen in research as well in applications for power forecasts in the field of renewable energy~\cite{Gen19} to consider the volatility of the weather forecasts, respectively. Probabilistic forecasts are especially useful for grid operators, as they allow them to devise optimal decisions under uncertainty considering expected costs and risks. In this context, it is desirable to take relations of renewable power plants into account by modeling dependencies and their combined effects on the electrical grid~\cite{Chen2018Bay}.

Modeling the relation, e.g., of neighbored farms, allows for decision making by taking simultaneous and time-shifted effects on the grid into account while improving the original forecast by making use of information from multiple farms.

Estimating these multivariate densities (taking spatial and temporal relation of farms into account) is still an open research problem and, e.g., addressed in~\cite{Tastu2015}.
Another important area of research for modeling temporal as well as spatial relations is scenario generation, for details we refer to~\cite{Chen2018Bay}. 
Similar to~\cite{Tastu2015}, we evaluate whether the QSNN is capable of modeling the relations between farms. 
In particular, we are interested in whether the QSNN is capable of modeling dependencies with increasing spatial distance between two farms for day-ahead forecasts. 
\begin{table}[h]
	\centering
	\noindent
	\begin{tabular}{N N N N}\toprule
		\multicolumn{1}{N }{\textbf{}} & \multicolumn{2}{c }{$\overline{\text{CRPS}}_{\text{\,DIR}}$ } & \multicolumn{1}{c }{Skill} \\  
		\cmidrule(lr){2-3}
		\cmidrule(ll){4-4}
		\multicolumn{1}{ N }{\textbf{Distance [\si{\km}]}} & \textbf{\centering Unconditional Gaussian} & \textbf{QSNN} & \textbf{QSNN} \\ 
		\cmidrule(lr){1-4}
		\multicolumn{1}{ c }{20} & 0.431 & 0.049 & 88.719 \\ 
		\multicolumn{1}{ c }{40} & 0.347 & 0.042 & 88.014 \\ 
		\multicolumn{1}{ c }{80} & 0.475 & 0.059 & 87.674 \\ 
		\multicolumn{1}{ c }{160} & 0.374 & 0.044 & 88.130 \\ 
		\multicolumn{1}{ c }{320} & 0.490 & 0.074 & 84.324\\ 
		\multicolumn{1}{ c }{720} & 0.342 & 0.044 & 87.112 \\ 
		\multicolumn{1}{ c }{1400} & 0.376 & 0.046 & 87.826  \\ 
		\bottomrule
	\end{tabular}
	\vskip 4mm
	\caption{Directional CRPS with respect to the distance of wind farms. Higher skills are better.}
	\label{tbl_crps_wind}
\end{table}


The temporal relation can be taken into account by utilizing important features~\cite{Jens2018}. 
The wind speed, as well as the wind direction, are shifted for two hours in time to take expected future as well as past events of the weather into account.
These additional features improve forecast quality. Further, this allows for modeling the temporal dependency between two farms. Spatial relation is modeled by the multivariate output of a regression model, i.e., 
we aim to forecast the power generation of multiple farms simultaneously.

As input to the QSNN, we use a dataset containing the power generation taken from 54 wind farms that are distributed throughout Europe~\cite{SBS19}. The dataset is referred to as \textit{EuropeanWindFarm} in the following.
The values were recorded hourly over two years (2016 and 2017). 
The generated power data were normalized by the maximum power generation of a farm. 

To evaluate the capability of the QSNN with respect to the distance between two farms, we select pairs of farms with a gap between $20 \si{\km}$ and $2800 \si{\km}$. As a baseline, we use an unconditional multivariate Gaussian model.
The pre-processing and model training are realized as follows: First, we split the recorded data from the year 2016 into training ($80\%$) and validation ($20\%$). The data from the year 2017 is solely used for testing.
Subsequently, we standardize the input features based on training data. In the next step, we train a linear regression model to issue deterministic central tendency estimates. Then, we train a QSNN with the Adam optimizer~\cite{KB15} for $10000$ epochs using a learning rate of $0.01$. We choose these parameters as they yielded reasonably good convergence of the training and validation loss. 
In this experiment, we utilize a neural network with two hidden layers and $100$ neurons in each layer and rectified linear unit activation functions to forecast the quantile levels $\tau \in \left\{ 0.01, 0.05, 0.1 , 0.2, \ldots , 0.9, 0.95, 0.99 \right\}$. 

The results depicted in Table~\ref{tbl_crps_wind} point out that the QSNN is capable of modeling the dependencies between two farms regardless of the evaluated distance. 
Only in the case of a distance of $320$~\si{\km} between wind farms, the $\overline{\text{CRPS}}_{\text{\,DIR}}$ increases to $0.074$, while still being about $84\%$ better than the baseline.
In most other cases the $\overline{\text{CRPS}}_{\text{\,DIR}}$ is below $0.05$, while the baseline is at least $0.342$ for all distances.
Also, for all other distances, the QSNN is more than $80$\% better compared to the baseline. 

\begin{figure*}[th]
	\centering
	\vskip -4mm
	\begin{subfigure}[t]{0.325\linewidth}
		\centering
		\includegraphics[width=1.\columnwidth]{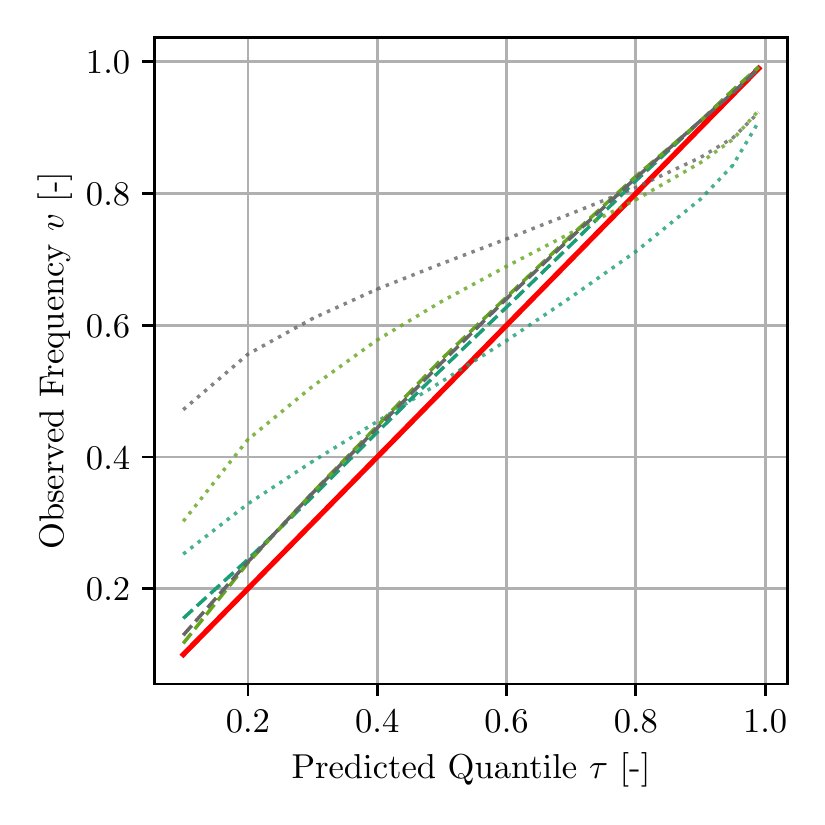}
		\vskip -2mm
		\caption{Q-Q-Plot for all cyclists trajectories}
		\label{fig:traj_qq_complete}
	\end{subfigure}
	\begin{subfigure}[t]{0.325\linewidth}
		\centering
		\includegraphics[width=\columnwidth]{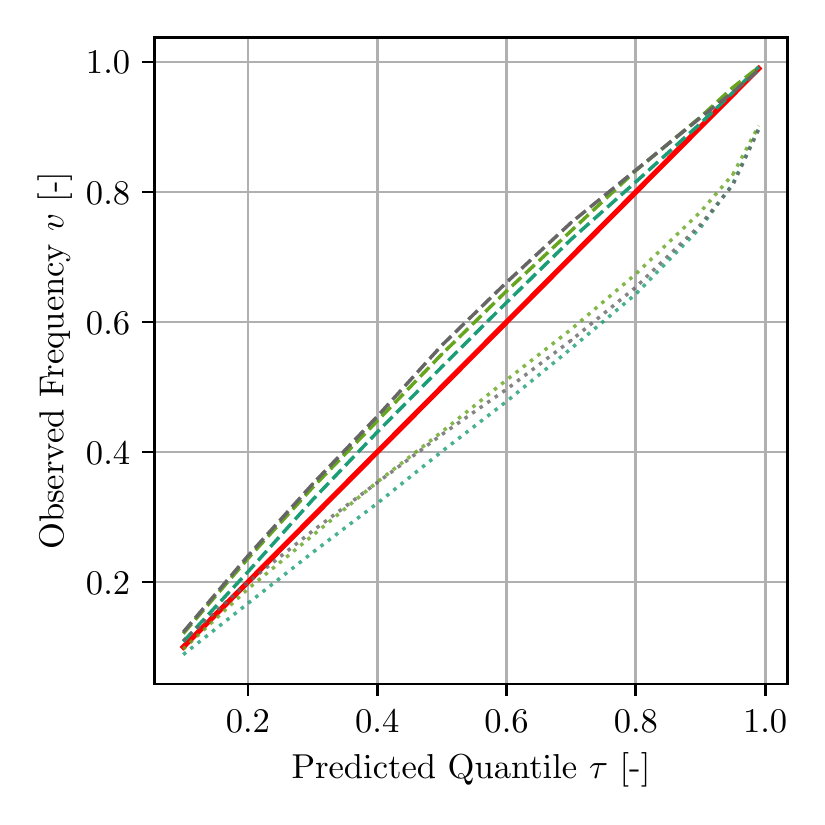}
		\vskip -2mm
		\caption{Q-Q-Plot for trajectories of moving cyclists.}
		\label{fig:traj_qq_moving}
	\end{subfigure}
	\begin{subfigure}[t]{0.325\linewidth}
		\centering
		\includegraphics[width=\columnwidth]{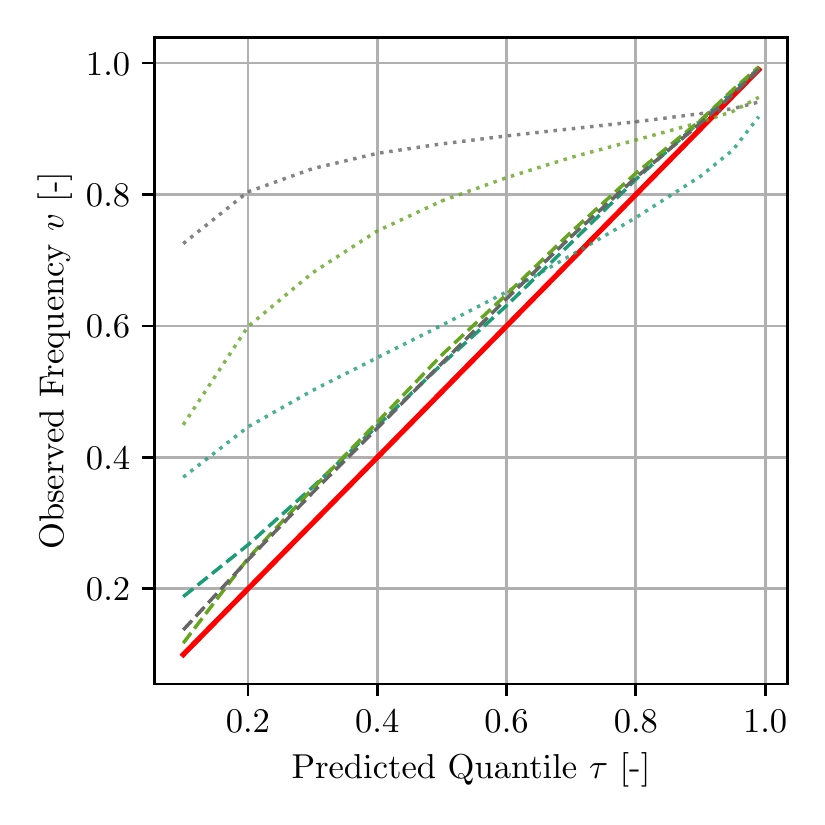}
		\vskip -2mm
		\caption{Q-Q-Plot for trajectories of waiting cyclists.}
		\label{fig:traj_qq_waiting}
	\end{subfigure}
	\vskip 5mm
	\caption{Q-Q-Plots of QRNN and conditional Gaussian approach for cyclist trajectory forecasting.
		Different forecasting horizons are compared: 0.2\,\si\second\ (cyan), 1.0\,\si\second\ (green), and 2.5\,\si\second\ (gray).
		The results of the QS estimates are depicted using dashed lines, while the results of the conditional Gaussian are represented using dotted lines. The red line corresponds to the theoretical optimal quantiles.
	}
	\label{fig:traj_qq_plots}
	\vskip -2mm
\end{figure*}

To get better insights regarding the modeling capabilities, we refer to Fig.~\ref{fig_wind_results_samples}. 
Fig.~\ref{fig_40km_sample} and Fig.~\ref{fig_1400km_sample} show examples of conditional forecasts, while Fig.~\ref{fig_wind_qq} shows their evaluation in terms of Q-Q-Plot.
The reliability for estimating the multivariate distribution shows considerably good results in Fig.~\ref{fig_wind_qq}.
Only for higher quantiles, there is a small deviation to the theoretical ones. Otherwise, the predicted quantiles are in-line with the expected quantiles showing reliable forecast quality of the QSNN.
This excellent reliability is partially surprising as the two forecasted farms are $1400$ \si \km\ distant, are located in a different terrain, and most likely have a weak correlation, i.e., they show different power outputs under similar weather conditions.

Even though the reference model was only unconditional, the results in terms of skill score and reliability plots show the superior performance of more than $80\%$ (cf. Table~\ref{tbl_crps_wind}) for modeling temporal and spatial relations between farms using the QSNN.

%% file: sections/results_cyclist_trajectory_forecasting.tex
The anticipation of other road users' behavior, i.e., the prediction of their future trajectories, is an important ingredient for safe autonomous driving. The behavior of other road users, especially vulnerable road users (VRU), is hardly predictable and only partly deterministic. Probabilistic forecasting in general and QS particularly are appealing for this task. In doing so, we model the future probability of residence, i.e., the predictive distribution of the VRU's future position. Using the sets computed by QS, we are able to derive statistical guarantees based on historical data, e.g., 
an area (i.e., QS) associated with 99\% probability of residence. These statistically derived bounds are an essential ingredient to make optimal decision under uncertainty, i.e., they are of great interest for safe trajectory planning~\cite{EBZ+17}. 
In this section, we apply our QS methodology to forecast the future short-term trajectories of cyclists. 
As in~\cite{GKD+16, BZD+17}, the goal is to issue trajectory predictions for a maximal forecasting horizon of 2.5\,\si\second.


For evaluation, we consider the data set consisting of $1056$ cyclist trajectories recorded at an urban intersection~\cite{VRUTraj18} in the city of Aschaffenburg, Germany. The data set covers different cyclist's motions, i.e., bending in, starting, and stopping maneuvers.
The cyclists were recorded using a wide-angle HD stereo camera system. The cameras are used to track the cyclists and to extract their head positions. The head position is triangulated, such that we obtain the 3D world position of the cyclist's head. We use the head as reference point to represent the cyclist's position. The trajectories are sampled with \SI{50}{\Hz}. We model the forecasting problem as a time series auto-regressive task, i.e., we consider features based on the past head trajectory, and we aim to predict the future course of the trajectory. The general forecasting methodology, including the features used and the choice of a location-independent ego-coordinate frame for prediction, is inspired by~\cite{EBZ+17, GKD+16, BZD+17}.

As in~\cite{GKD+16}, we consider the past position and velocity of the cyclist in the ground plane as well as the yaw rate.
We approximate these trajectories using polynomials based on an orthogonal polynomial basis in a sliding window manner. The coefficients of the basis polynomials represent best estimators (in the least-squares sense) of the slope, curvature, etc. of the approximated time series. We use these features as input for our probabilistic forecasting methodology.
We use a sliding window length of \SI{2}{\second} and polynomial coefficients up to the third degree derived from longitudinal and lateral position in a location-independent ego-coordinate frame. We also consider features derived from the coefficients of the orthogonal basis polynomials fitted over the past cyclist's yaw rate. Additionally, we use the current approximation error of the polynomial as input features. 
Moreover, we consider the currently predicted longitudinal and lateral movement state of the cyclists as input~\cite{BRZ+17}. The longitudinal movement state is characterized by waiting, starting, moving, or stopping, while the lateral movement state is characterized by moving straight, turning right, or turning left. In total, we use $27$ features.

First, we standardize the input features in a preprocessing step. 
We consider the forecast lead times $t \in \left( 0.2\,\si\second, 0.5\,\si\second, 1.0\,\si\second, 1.5\,\si\second, 2.0\,\si\second, 2.5\,\si\second \right)$. The deterministic point estimate is realized by fitting a neural network. This network is referred to as the deterministic point forecasting network in the following. Subsequently, we apply our QS approach. Therefore, we fit a QSNN comprising one hidden layer with ten neurons. We fit a separate QSNN model for each lead time. As activation function, we use a hyperbolic tangent. The output neuron is a simple linear unit. We consider the quantile levels $\tau \in \left\{ 0.1, 0.2, \ldots , 0.9, 0.95, 0.99 \right\}$.
We train the network for $5000$ epochs using stochastic gradient descent and the Adam optimizer~\cite{KB15}. 
We consider 819 cyclist trajectories to train the QSNN, while 237 trajectories are used for testing. 
We compare our approach to a neural network with multivariate Gaussian output, as detailed in Section~\ref{subsec:multivariate_gaussian_baseline}. However, we use the neural network only to predict the covariance, i.e., similar to the QSNN, we examine forecast-adjusted observations (cf. Eq.~\ref{eq:forecast-adjusted_predictions}).
Note that this approach for probabilistic trajectory forecasting is similar to the approach by Zernetsch~et~al.~\cite{ZRK+19}, which can be considered state-of-the-art. The architecture of the network is the same as the one we used for the QSNN, i.e., one hidden layer with ten neurons and the hyperbolic tangent as activation function. For the sake of brevity, this is referred to as the conditional Gaussian approach in the following. As baseline model, we consider an unconditional multivariate Gaussian model. We fit an unconditional Gaussian covariance 
using maximum likelihood estimation to the outputs of the deterministic point estimate neural network.

\begin{figure}[b]
	\centering
	\includegraphics[width=0.35\textwidth, clip, trim=0 10 0 10]{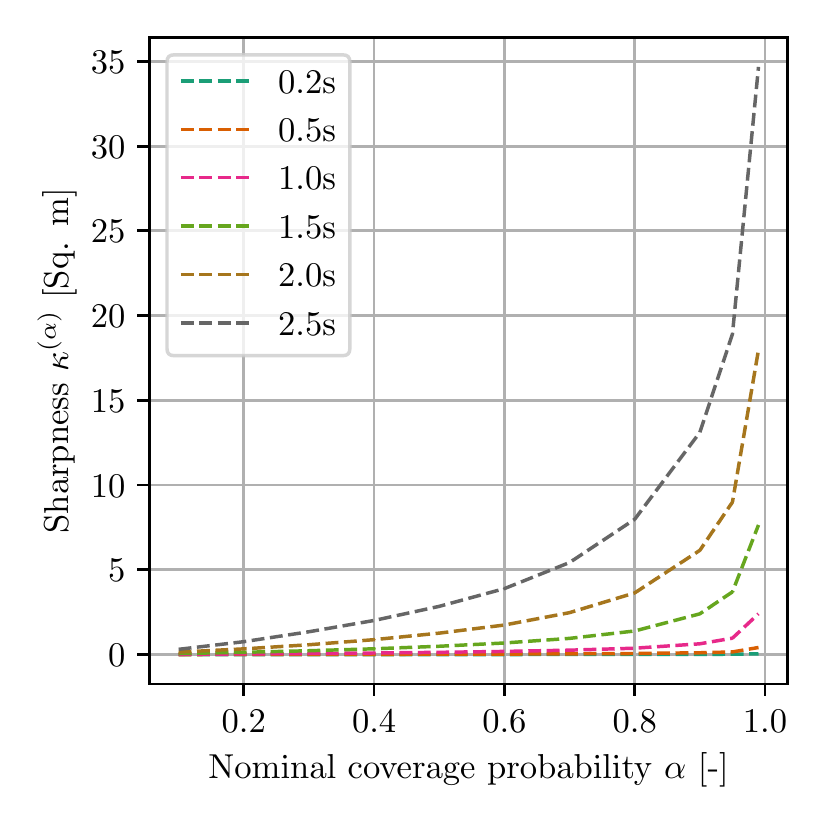}
	\vspace{2em}
	\caption{Sharpness diagram showing evaluation of QS predictions of future cyclists' trajectories for different forecasting horizons.}
	\label{fig:traj_sharpness}
\end{figure}

To evaluate the reliability of our novel QS forecasting technique, we evaluate the reliability of the issued forecasts using a
Q-Q-plot. The results obtained on the complete data set as well as examinations according to different movement types are depicted in Fig.~\ref{fig:traj_qq_plots}. As a reference, we also depict the results of the conditional Gaussian approach. 
As it can be seen in Fig.~\ref{fig:traj_qq_complete}, the QS issues well-calibrated forecasts for the entire range of quantile levels, while the conditional Gaussian approach generally issues underconfident probability estimates. 
Moreover, inspecting the dependence on the forecasting horizon, we observe that the QS approach also issues well-calibrated forecasts even for higher forecasting horizons, while the conditional Gaussian approach deteriorates. 

Next, we examine the sharpness of the issued predictive distributions. Therefore, we evaluate the average size of the covered area with respect to different nominal coverage probabilities. The results for different forecasting horizons are depicted in Fig.~\ref{fig:traj_sharpness}. As one might expect, we can observe an extreme increase in the average sharpness (i.e., area covered) 
when reaching the extreme quantiles. We can also see that for shorter forecasting horizons up to 1.5\,s 
even the extreme coverage probabilities ($\alpha \geq 0.95$) levels are relatively small. For larger horizons, we observe an exponential increase.

\begin{figure}[h]
	\centering
	\includegraphics[width=0.35\textwidth, clip, trim=0 10 0 10]{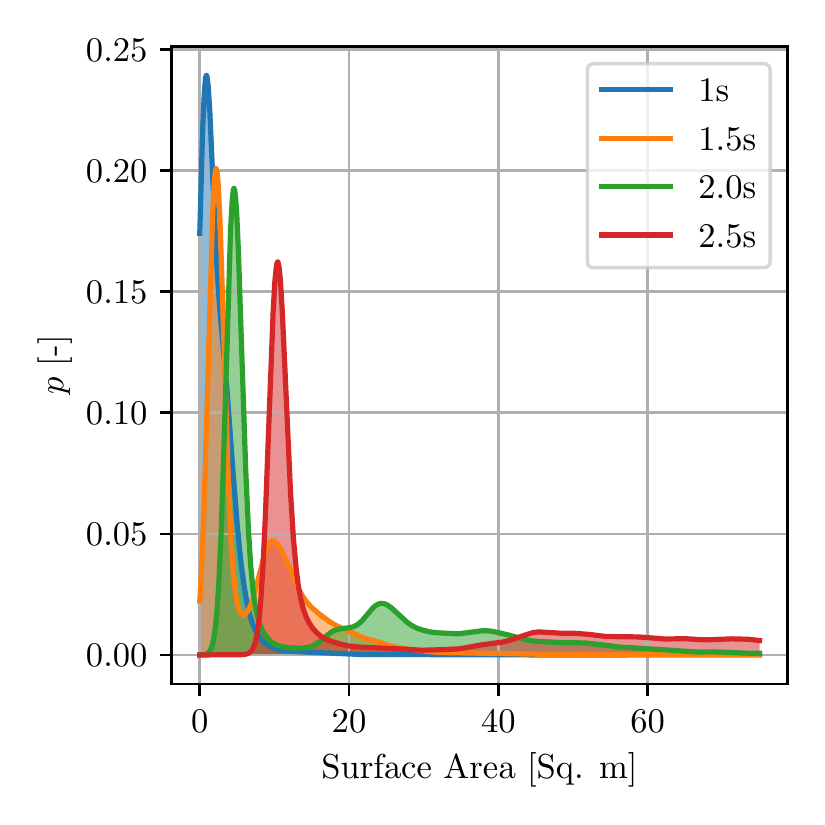}
	\vspace{2em}
	\caption{Distribution of surface area covered by the $\tau=0.99$-QS prediction for different forecasting horizons}
	\label{fig:traj_sharpness_dist}
\end{figure}

However, the mean coverage does not tell us anything about the distribution of forecasted QS. We cannot deduce whether our method perhaps always only predicts the same predictive distribution regardless of the current input.
Therefore, we inspect the distribution of the surface area covered by a distinct coverage probability level $\alpha$.
We considered a probability coverage of $\alpha = 0.99$. The surface area distributions are depicted in Fig.~\ref{fig:traj_sharpness_dist}.
Here, we also observe sharp distributions concerning the surface area for shorter forecasting horizons, while for larger horizons, these distributions are wider and more skewed. 
This indicates that the QSNN is capable of automatically identifying situations where it can predict the future cyclist position with smaller or higher uncertainty.

The results of the average directional CRPS for different forecasting horizons are depicted in Table~\ref{tab:traj_crps_skill}.
As expected, we observe that our forecasting methodology outperforms the baseline with an improvement of up to $99\%$ with respect to the average directional CRPS.

\begin{table}[h]
    \vskip -1mm
	\begin{center}
		\begin{tabular}{  c  c  c  c  c  c  c  }
			\toprule
			Lead Time & 0.2\,s & 0.5\,s & 1.0\,s & 1.5\,s & 2.0\,s & 2.5\,s \\ 
			\cmidrule{1-7}
			$\overline{\text{CRPS}}_{\text{DIR}}$ [m] & 0.008 & 0.022 & 0.053 & 0.103 & 0.164 & 0.242 \\ 
			Skill [\%] & 99 & 97 & 96 & 93 & 89 & 83 \\ 
			\bottomrule
		\end{tabular}
	\end{center}
	\vskip 4mm
	\caption{$\overline{\text{CRPS}}_{\text{DIR}}$ and skill of QS forecasting methodology with respect to baseline, i.e., unconditional Gaussian model.}
	\label{tab:traj_crps_skill}
\end{table}
%

%% file: sections/conclusion.tex
\section{\large Conclusions and Future Work}
\label{sec_conclusion}
In this article, we present a generalization of quantile regression for multivariate targets, which allows for flexible non-linear, non-parametric, probabilistic forecasts. Our QS approach consists of a two-stage model involving a deterministic point forecasting followed by the actual probabilistic model using the novel QS concept. 
The key innovation is the directional representation of the data points through direction and vector length.
Fitting a one-dimensional quantile regression model on the vector length with the direction as an additional input, our QSNN approach is able to represent star domain sets. Using this transformation and the two-stage model, we can easily extend arbitrary existing deterministic central tendency forecasts with QSNN to fully probabilistic forecasting models.
Yet, to evaluate our novel approach and to compare it to other existing multivariate probabilistic forecasting techniques, we present methods on how to assess the reliability, sharpness, and skill (i.e., the directional CRPS) of the forecasts issued by our QS approach.
We show the applicability of our approach on synthetic data and two-real world datasets. 
The latter involves the modeling of dependencies in the power generation between wind parks. 
Moreover, the QS methodology improves the state-of-the-art performance for short-term trajectory forecasting of cyclists. 

In our future work, we will work on extending the QS approach to model and represent multimodal predictive distributions. 
Currently, QS only models aleatoric uncertainty. Yet, in many cases (e.g., active learning) also modeling epistemic uncertainties is required.
Therefore, in our future work, we will consider the combination of the QS paradigm with Bayesian neural networks~\cite{Bis06}.
Moreover, we will also investigate the use of different machine learning models such as gradient boosted regression trees for predicting QS. Another interesting direction for future research is the integration of trigonometric basis functions into the output of the neural network to constrain the flexibility of the QS. Constrain the flexibility might potentially lead to smoother surfaces and a better generalization of the model. Using fixed basis functions to represent the model output is also appealing when it comes to representing QS in high-dimensional output spaces, i.e., avoiding the computationally inefficient sampled-based representation. Moreover, also the joint optimization of models used for the deterministic point forecast and prediction of QS is an exciting direction of future research.

